\documentclass[twocolumn,american,pra]{revtex4-1}
\usepackage[T1]{fontenc}
\pdfoutput=1 
\usepackage[latin9]{inputenc}
\usepackage{amsmath}
\usepackage{amssymb}
\usepackage{graphicx}



\usepackage{babel}
\begin{document}

\title{Nonlinear plasmonic amplification via dissipative soliplasmons }

\author{Albert Ferrando}

\affiliation{Departament d'Òptica. Interdisciplinary Modeling Group, \emph{InterTech.} Universitat
de València. Spain}
\begin{abstract}
In this contribution we introduce a new strategy for the compensation
of plasmonic losses based on a recently proposed nonlinear mechanism:
the resonant interaction between surface plasmon polaritons and spatial
solitons propagating in parallel along a metal/dielectric/Kerr structure.
This mechanism naturally leads to the generation of a quasi-particle
excitation, the so-called soliplasmon resonance. We analyze the role
played by the effective nonlinear coupling inherent to this system
and how this can be used to provide a new mechanism of quasi-resonant
nonlinear excitation of surface plasmon polaritons. We will pay particular
attention to the introduction of asymmetric linear gain in the Kerr
medium. The unique combination of nonlinear propagation, nonlinear
coupling and gain give rise to a new scenario for the excitation of
long- range surface plasmon polaritons with distinguishing characteristics.
The connection between plasmonic losses and soliplasmon resonances
in the presence of gain will be discussed.
\end{abstract}
\maketitle

\section{Introduction}

High intensity electromagnetic fields are characteristic for a surface
plasmon polariton (SPP) in metal/dielectric interfaces \cite{maier07}.
These high intensity fields are achieved nearby the metal, so that
this region becomes a natural scenario for the presence of nonlinear
effects. If the interface is constituted by a metal and a Kerr medium,
the characteristics of the propagation of the SPP can be strongly
modified by the presence of the nonlinearity. In such a case, the
propagation constant of the SPP becomes dependent on the plasmon amplitude.
These nonlinear effects on SPPs were studied long ago defining what
it is known as a nonlinear plasmon \cite{Agranovich1980,Mihalache1984a,Stegeman1985a,Ariyasu1985a}.
The high intensities in the vicinity of the metal are also responsible
of a variety of nonlinear effects. When the confinement of the electromagnetic
field takes place at a nanometric scale, as in plasmonic nanostructures,
a panoply of these nonlinear effects are expected to be strongly enhanced
\cite{Boardman2011,Kauranen2012a,Zayats2013,Shadrivov2015}. 

The high value of SPP losses for a single metal/dielectric interface
in the visible and infrared domains dramatically reduces the SPP propagation
length to a few tens of microns, a value clearly insufficient for
many practical applications. Strategies to increase the SPP propagation
length are then necessary to overcome this important limitation (see
\cite{Leosson2012a} and references therein.) The use of more complex
waveguiding geometries than a simple interface permits to access to
SPP modes with different spatial distributions. Some of these modes,
those in which the mode amplitude is more localized in the dielectric
than in the metal, present larger propagation lengths \cite{Oulton2008}.
This is the case of the long-range surface plasmon polariton (LRSPP),
the low-loss symmetric mode of a thin metal slab immersed between
dielectrics with similar refractive indices \cite{Berini2009,Holmgaard2010,Volkov2011}.
This strategy can increase the propagation length up to a few centimeters.
In order to compensate plasmonic losses even further, the addition
of gain into the dielectric part of the waveguide has been a second
step exploited in recent years by different groups\cite{Nezhad2004a,Noginov2008a,Ambati2008,Grandidier2009,Leon2010,Gather2010,Zhang2011,Garcia-Blanco2011,Gao2012}.
In the previous context, the combination of nonlinear effects with
gain in plasmonic waveguides appears as a natural consequence. Different
theoretical proposals of gain-assisted nonlinear waveguiding structures
have been developed recently \cite{Marini2011a,Argyropoulos2013,Adelpour2014,Ding2014a,Malomed14,Marini2015,Wang2016}.
Theoretical tools to analyze periodic plasmonic waveguides and metamaterials
with loss and gain have been also reported \cite{Sukhorukov2014a}.
Our contribution in this article points out in the same direction
although it presents important distinguishing features.

Our approach is based on the concept of soliplasmon resonance. A
soliplasmon resonance can be understood as a quasiparticle combining
a SPP mode with a spatial soliton as a result of its resonant or quasi-resonant
interaction during propagation along a metal/dielectric/Kerr (MDK)
interface \cite{Bliokh2009,Milian2012a}. Although the plasmon-soliton
term refers generically to hybrid nonlinear solutions involving metal
and Kerr media and they were already studied during the 1980s \cite{Agranovich1980,Mihalache1984a,Stegeman1985a,Ariyasu1985a},
they do not necessarily deal with situations in which they exhibit
a manifest resonant behavior between plasmon and soliton \cite{Feigenbaum2007a,Davoyan2009,Marini2010,Ye2010a,Marini2011a,Walasik2012,Ginzburg2013,Walasik2014,Sadeghi2016}. 

Mathematically, a plasmon-soliton solution of the nonlinear Maxwell's
equations is formalized as a soliplasmon when it can be approximated
by the so-called soliplasmon \emph{ansatz }\cite{Ferrando2013b}\emph{.
}Physically, the soliplasmon \emph{ansatz }represents a variational
solution of nonlinear Maxwell's equations in which the plasmon and
soliton components are distinguishable \textemdash thus, spatially
separated\textemdash{} and not strongly overlapping. The variational
equations for the soliplasmon ansatz are rather simple in form since
they correspond to two coupled oscillators: one linear (for the SPP
variational amplitude) and the other nonlinear (for the soliton amplitude).
However, the coupling presents a distinguishing feature: it is nonlinear
and evanescent with the soliton amplitude $\left|C_{s}\right|$ and
position $a$, i.e., $q\propto\exp\left(-K\left|C_{s}\right|a\right)$
\cite{Bliokh2009,Ferrando2013b}. Besides, it is non-symmetric, which
means that the soliton-to-plasmon coupling coefficient $q$ is generically
much larger than the plasmon-to-soliton one $\bar{q}$ \cite{Ferrando2013b}.
Although obtained explicitly for a MDK interface, the variational
soliplasmon approach applies to more general 2D waveguiding structures
in which the linear component can be a plasmonic mode (not just a
SPP) and the nonlinear component can represent a nonlinear dielectric
mode (not just a spatial soliton of an homogeneous medium.) The particular
characteristics of this nonlinear evanescent coupling makes the soliplasmon
model different from previous approaches for similar metal/Kerr systems
in which coupling terms turns out to be linear and symmetric \cite{Ye2010a,Marini2010}.
The distinguishing properties of the nonlinear evanescent coupling
was reported in \cite{Eksioglu2011a,Eksioglu2012,Eksioglu2013a},
in which the soliplasmon model \textemdash for \emph{symmetric }coupling\textemdash{}
was mapped into that of a bosonic Josephson junction (BJJ). The dynamics
of this nonlinear BJJ was proven to be \textemdash qualitatively and
quantitatively\textemdash{} different from that of a standard linear
coupling BJJ, thus indicating the crucial role played by the nonlinear
evanescent coupling.

In this article we will analyze the profound effect that the addition
of gain in the Kerr medium has in the propagation properties of the
plasmonic component of a soliplasmon resonance. The content of the
article is distributed as follows: in Section \ref{sec:Coupled-oscillator-model}
we introduce the dissipative soliplasmon model; in Section \ref{sec:Spin-model}
we construct the spin model associated to this model; in Sections
\ref{sec:Stationary-spin-states} and \ref{sec:Stationary-dissipative-soliplasm}
we obtain the stationary dissipative soliplasmon solutions of the
spin model; in Section \ref{sec:Stationary-critical-gain} we present
the phenomenon of critical gain linked to these solutions; and, finally,
in Section \ref{sec:Nonlinear-amplification} we unveil how dissipative
soliplasmons can act as mediators for strong nonlinear plasmonic amplification.

\begin{figure*}
\includegraphics[width=1\textwidth]{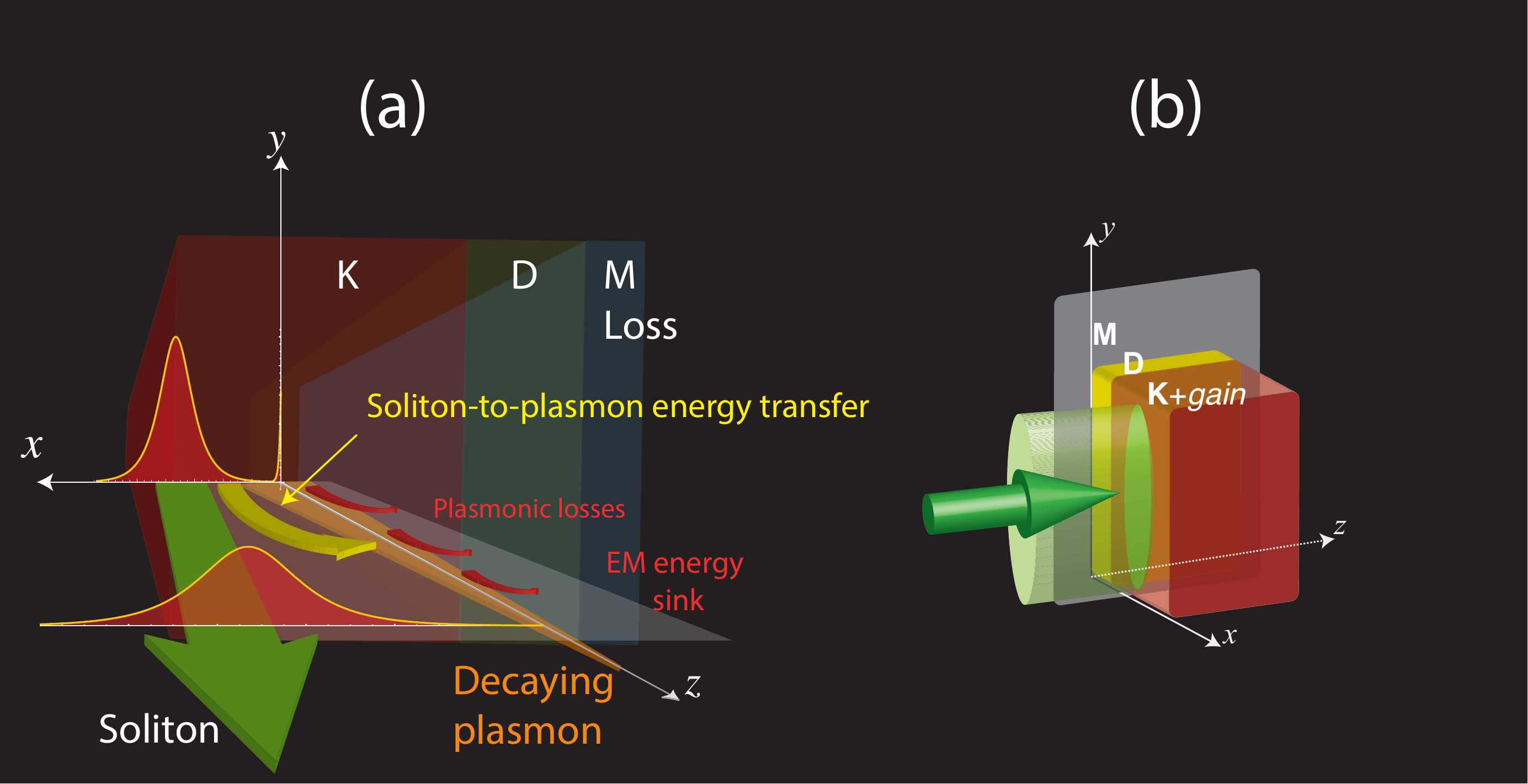}

\caption{(a) Scheme of the flux of EM energy in a soliplasmon structure with
just metal losses. The SPP component is localized transversely ($x$-axis)
with a maximum at the MD interface ($x=0)$ whereas the spatial soliton
component is centered at a certain distance $x=a$ from the MD interface
within the Kerr medium. Both propagate in parallel to the MD and DK
interfaces. (b) Characteristic MDK structure supporting dissipative
soliplasmons in which gain is located in the Kerr medium in order
to compensate for metal losses. \label{fig:MDK_energy_flux}}
\end{figure*}

\section{Soliplasmon model with loss and gain\label{sec:Coupled-oscillator-model}}

The variational model for the non-dissipative soliplasmon system was
developed with detail in Ref.\cite{Ferrando2013b}. A summary of its
main results is given in the Appendix A. The variational model for
conservative soliplasmons was obtained assuming that the propagation
of a monochromatic quasi-stationary TM solution along a MDK structure
such as in Fig.~\ref{fig:MDK_energy_flux} can be approximated by
the soliplasmon \emph{ansatz:
\begin{equation}
E_{x}(x,z)=C_{p}(z)e_{p}(x)+C_{s}(z)\mathrm{sech}\left[\sqrt{\frac{\gamma}{2}}\left|C_{s}(z)\right|(x-a)\right],\label{eq:variational_ansatz-1}
\end{equation}
}where $e_{p}(x)$ is a linear SPP solution of the MD interface, $C_{p}(z)$
and $C_{s}(z)$ are the two variational parameters, $a$ is the soliton
position and $\gamma=\left(3/4\right)\varepsilon_{0}cn_{2}$ is the
nonlinear coefficient of the Kerr medium. The other components of
the TM mode, i.e., $H_{y}$ and $E_{z}$, can be obtained by using
additional constraints valid under reasonable approximations \cite{Ferrando2013b}.
In this first approximation the metal is considered to be an ideal
conductor with $\mathrm{Im}(\varepsilon_{m})=0$. However, for realistic
lossy metals in which $\mathrm{Im}(\varepsilon_{m})\neq0$ the plasmon
propagation constant becomes complex $\beta_{p}^{2}=\beta'{}_{p}^{2}+i\beta''{}_{p}^{2}$.
Consequently, the paraxial plasmon propagation constant also becomes
complex $\mu_{p}=\mu'_{p}+i\mu''_{p}$. 

Besides the effect of metal losses generating the complex SPP plasmon
propagation constant, it is possible to add lineal gain or loss in
the Maxwell's equation description of the MDK structure in Fig.\,\ref{fig:MDK_energy_flux}(a).
This can be done mathematically by considering a modified complex
permittivity function $\widetilde{\varepsilon}_{L}(x)=\varepsilon_{L}(x)+i\delta\varepsilon_{L}(x)$,
where $\varepsilon_{L}(x)$ is the linear permittivity function of
the MDK structure (in which we assume that metal losses are already
included) and the imaginary part $\delta\varepsilon_{L}(x)$ provides
the loss/gain profile introduced additionally in the structure by
means of other lossy or active components. It is possible to incorporate
into our variational model these gain/loss effects in a relatively
simple manner if we assume that their transverse distribution is clearly
spatially separated between the plasmon and soliton regions with small
overlapping. In this way, additional loss/gain occurring in the SPP
localization region, i.e., in the MD layers, would be given by a modified
complex profile $\widetilde{\varepsilon}_{p}(x)=\varepsilon{}_{p}(x)+i\delta\varepsilon{}_{p}(x)$,
where $\varepsilon{}_{p}(x)$ would be the permittivity profile of
the MD structure (including metal losses) given by the standard expression
in Eq.(\ref{eq:permittivity_MD}) with complex $\varepsilon_{m}$
. On the other hand, loss/gain in the soliton localization region
would be analogously described by the modified permittivity function
$\widetilde{\varepsilon}_{s}(x)=\varepsilon{}_{s}(x)+i\delta\varepsilon{}_{s}(x)$,
defined in the Kerr medium. Taking the imaginary part $\delta\varepsilon{}_{L}(x)$
as a small perturbation of the unperturbed one $\varepsilon{}_{L}(x)$
permits to calculate easily the first-order correction \textemdash which
will be purely imaginary\textemdash{} in terms of the unperturbed
values using standard perturbation theory. The first-order corrections
to the propagation constants are given by \footnote{$N_{p}$ is the plasmon norm of the unperturbed system, which is complex
in the case of a lossy metal (see Eq.(\ref{eq:norms}).) However,
its imaginary part is smaller than the real part and $N''_{p}=\mathrm{Im}\left(N_{p}\right)\rightarrow0$
when $\mathrm{Im}\left(\varepsilon_{m}\right)\rightarrow0$ so $N''_{p}$
can be considered a perturbation. Since $\delta\varepsilon{}_{p}$
is itself a perturbation, we can approximate $N_{p}$ by its real
part at first order. }:
\begin{align}
\delta\beta''{}_{p}^{2} & =\frac{1}{N_{p}}\int_{\mathbb{R}}\!\!dx\delta\varepsilon{}_{L}(x)f_{p}^{2}(x)\nonumber \\
 & \approx\frac{1}{N_{p}}\int_{\Omega_{\mathrm{MD}}}\!\!dx\delta\varepsilon{}_{p}(x)f_{p}^{2}(x)\\
\delta\beta''{}_{s}^{2} & =\frac{1}{N_{s}}\int_{\mathbb{R}}\!\!dx\delta\varepsilon{}_{L}(x)\mathrm{sech^{2}}\left[\sqrt{\frac{\gamma}{2}}\left|C_{s}\right|(x-a)\right]\label{eq:integrals_perturbed_beta2}\\
 & \approx\frac{1}{N_{s}}\int_{\Omega_{K}}\!\!dx\delta\varepsilon{}_{s}(x)\mathrm{sech^{2}}\left[\sqrt{\frac{\gamma}{2}}\left|C_{s}\right|(x-a)\right],
\end{align}
where we have used the spatial localization assumption on the loss/gain
functions to determine the integration domains $\Omega_{\mathrm{MD}}$
($x\le d$) and $\Omega_{K}$ ($x>d$) \cite{Ferrando2013b}. So that,
the full propagation constants including the metal loss and the contribution
introduced by extra gain or loss sources in the system are: 
\begin{align}
\beta_{p}^{2} & =\beta'{}_{p}^{2}+i\left(\beta''{}_{p}^{2}+\delta\beta''{}_{p}^{2}\right)\nonumber \\
\beta_{s}^{2} & =\beta'{}_{s}^{2}+i\delta\beta''{}_{s}^{2}.\label{eq:complex_prop_const}
\end{align}
The calculation of the complex \emph{paraxial} propagation constants
\begin{align*}
\mu_{p} & =\mu'_{p}+i\mu''_{p}\\
\mu_{s} & =\mu'_{s}+i\mu''_{s}
\end{align*}
is performed by means of Eqs.(\ref{eq:paraxial_prop_const}) but using
now the complex propagation constants (\ref{eq:complex_prop_const}).
The real part of the soliton paraxial propagation constant depends
nonlinearly on the soliton component $C_{s}$:
\[
\mu'_{s}\left(C_{s}\right)=\frac{k_{0}\gamma}{4\varepsilon_{\mathrm{K}}^{1/2}}|C_{s}|^{2},
\]
where the nonlinear coefficient $\gamma$ is given by $\gamma=\left(3/4\right)\varepsilon_{0}cn_{2}$,
$n_{2}$ being the nonlinear index of the Kerr medium and $\varepsilon_{K}$
its real linear dielectric constant.

An analogous analysis holds also for the coupling coefficients $q$
and $\bar{q}$. The local permittivity functions $\triangle\varepsilon_{p}$
and $\triangle\varepsilon_{s}$, as defined in Eqs.(\ref{eq:local_permittivity}),
pick up additional imaginary parts from the loss/gain function $\delta\varepsilon_{L}$
associated to the new total permittivity function $\widetilde{\varepsilon}_{L}(x)=\varepsilon_{L}(x)+i\delta\varepsilon_{L}(x)$.
According to their definitions, the perturbed local permittivity functions
are now:
\begin{align*}
\triangle\widetilde{\varepsilon}_{p}(x) & =\widetilde{\varepsilon}_{L}(x)-\varepsilon_{p}(x)=\triangle\varepsilon{}_{p}(x)+i\delta\varepsilon_{L}(x)\\
\triangle\widetilde{\varepsilon}_{s}(x) & =\widetilde{\varepsilon}_{L}(x)-\varepsilon_{s}(x)=\triangle\varepsilon{}_{s}(x)+i\delta\varepsilon_{L}(x),
\end{align*}
where $\triangle\varepsilon{}_{p}$ and $\triangle\varepsilon_{s}$
would be given by Eqs.(\ref{eq:local_permittivity}) and correspond
to an unperturbed MDK structure in which only $\varepsilon_{m}$ is
complex (lossy metal). In this way, the first-order correction to
the coupling coefficients will be given by the overlapping integrals:
\begin{align}
\delta q & =\frac{k_{0}}{2\varepsilon_{K}^{1/2}N_{p}}\int_{\mathbb{R}}\!\!dxf_{p}(x)\delta\varepsilon{}_{L}(x)\mathrm{sech}\left[\sqrt{\frac{\gamma}{2}}\left|C_{s}\right|(x-a)\right]\nonumber \\
\delta\overline{q} & =\frac{k_{0}}{2\varepsilon_{K}^{1/2}N_{s}}\int_{\mathbb{R}}\!\!dxf_{p}(x)\delta\varepsilon{}_{L}(x)\mathrm{sech}\left[\sqrt{\frac{\gamma}{2}}\left|C_{s}\right|(x-a)\right],\label{eq:perturbed_couplings}
\end{align}
in such a way the whole coupling coefficients take the form:

\begin{align*}
\widetilde{q} & =q+i\delta q=q'+i\delta q''\\
\widetilde{\overline{q}} & =\bar{q}+i\delta\bar{q}=\bar{q}'+i\delta\bar{q}'',
\end{align*}
where $q'=\mathrm{Re}\left(\widetilde{q}\right)$ and $\delta q''=\mathrm{Im}\left(\widetilde{q}\right)$
and analogously for $\widetilde{\overline{q}}$. Note that both $q$
and $\delta q$ are complex so that they are not exactly the real
and imaginary parts of $\widetilde{q}$. However, their imaginary
parts are relatively small so that it is not a bad approximation to
consider $q\approx q'$ and $\delta q\approx\delta q''$.

In summary, the introduction of localized loss or gain in the MDK
structure can be also analyzed using the same soliplasmon model in
Appendix \ref{sec:Variational-equations} but with complex propagation
constants and couplings:
\begin{align}
-i\frac{dC_{p}}{dz} & =\left(\mu'_{p}+i\mu''_{p}\right)C_{p}+\left(q'+i\delta q''\right)C_{s}\nonumber \\
-i\frac{dC_{s}}{dz} & =\left(\mu'_{s}+i\mu'_{s}\right)C_{s}+\left(\bar{q}'+i\delta\bar{q}''\right)C_{p},\label{eq:NL_oscillator_loss_gain}
\end{align}
We will refer to this model as the \emph{dissipative} soliplasmon
model.

\section{Spin model \label{sec:Spin-model}}

In this article, we are interested in the search for solutions in
which the intrinsic plasmon loss is compensated by the gain in the
Kerr medium. In the absence of coupling the soliplasmon model equations
(\ref{eq:NL_oscillator_loss_gain}) indicate that the plasmon and
soliton parameters evolve as $C_{p}(z)\sim e^{i\mu'_{p}z}e^{-\mu''_{p}z}$
and $C_{s}(z)\sim e^{i\mu'_{s}z}e^{-\mu''_{s}z}$, respectively. Thus,
in order to make the loss/gain balance more explicit we write the
imaginary parts of the paraxial propagations constants as $\mu''_{p}=k_{0}l$
and $\mu''_{p}=-k_{0}g$, where $l$ and $g$ are the positive definite
dimensionless loss and gain coefficients. Note that due to the lack
of symmetry of the MDK structure these effective coefficients correspond
to an asymmetric spatial distribution of loss and gain, thus it is
expected that $l\neq g$ in the most general case. If we divide Eqs.(\ref{eq:NL_oscillator_loss_gain})
by $k_{0}$, all the coefficients of the system of differential equations
becomes dimensionless. Besides, if we introduce the new propagation
variable $\bar{z}=k_{0}z$, the differential operator becomes dimensionless
as well. This procedure is equivalent to set $k_{0}=1$, so that $\mu_{p}$,
$\mu_{s}$, $q$ and $\overline{q}$ become dimensionless and, in
addition, the measurement of axial distances is done in units of $k_{0}^{-1}=\lambda/2\pi$
(from now on we maintain the notation of $z$ as the axial variable
with this interpretation).

The system of coupled equations (\ref{eq:NL_oscillator_loss_gain})
admits the following matrix representation:
\begin{equation}
-i\frac{dC}{dz}=HC,\label{eq:evol_matrix_representation}
\end{equation}
where $C=\left(C_{p},C_{s}\right)^{\top}$ and 
\begin{equation}
H=\left[\begin{array}{cc}
\mu_{p} & q\\
\bar{q} & \mu_{s}
\end{array}\right]\label{eq:hamiltonian}
\end{equation}
is the Hamiltonian of the system, where, in principle, all its elements
are complex. 

The matrix $H$ is non-Hermitian ($H\neq H^{\dagger}$, $H^{\dagger}$
being the conjugate transpose or adjoint matrix of $H$) but it admits
the following decomposition in Hermitian matrices:
\[
H=\Pi+i\Sigma,
\]
where
\begin{align}
\Pi & =\frac{H+H^{\text{\ensuremath{\dagger}}}}{2}=\left[\begin{array}{cc}
\mu'_{p} & q_{0}\\
q_{0}^{*} & \mu'_{s}
\end{array}\right]\nonumber \\
\Sigma & =\frac{-i}{2}\left(H-H^{\text{\ensuremath{\dagger}}}\right)=\left[\begin{array}{cc}
l & -i\Delta_{0}\\
i\Delta_{0}^{*} & -g
\end{array}\right],\label{eq:pi_sigma_decompositoin}
\end{align}
in which we have introduced the notation $q_{0}\equiv\frac{1}{2}\left(q+\overline{q}^{*}\right)$
and $\Delta_{0}\equiv\frac{1}{2}\left(q-\overline{q}^{*}\right)$.
We see that $\Pi$ represents the Hermitian part of the Hamiltonian
$H.$ If $\Sigma$ were absent, the evolution of the system would
be conservative \textendash energy would be conserved\textemdash{}
and the norm of the vector $C$ would be preserved since the evolution
operator $\exp\left(iHz\right)$ would be unitary. The matrix $\Sigma$
is responsible of the breaking of the hermiticity of $H$ and, thus,
of the unitary evolution of $C$.

Now we introduce the equivalent of the density matrix of a pure state
$\rho=\left|C\right\rangle \left\langle C\right|$, which in matrix
notation would be simply $\rho=C\cdot C^{\dagger}$, where the dot
represents standard matrix multiplication. Our goal is to find the
evolution equation for $\rho$ taking into account that $H$ is no
longer Hermitian. In order to do so we need first to find the adjoint
matrix equation associated to (\ref{eq:evol_matrix_representation}),
which is simply
\[
i\frac{dC^{\dagger}}{dz}=C^{\dagger}H^{\dagger}.
\]
Using both equations it is straightforward to prove that:
\begin{align*}
-i\frac{d\rho}{dz} & =H\rho-\rho H^{\dagger}=\left[\Pi,\rho\right]+i\left\{ \Sigma,\rho\right\} ,
\end{align*}
where $\left[A,B\right]=AB-BA$ and $\left\{ A,B\right\} =AB+BA$
are the commutator and anticommutator matrices, respectively. In order
to convert the previous equation into an evolution equation for a
spin model we resort to the standard decomposition of the $\rho$
matrix in terms of the identity and Pauli matrices $\left\{ \tau_{0},\tau_{1},\tau_{2},\tau_{3}\right\} $
(where $\tau_{0}=I$), which constitute a basis of 2D matrices :
\[
\rho=\sum_{i=0}^{3}S_{i}\left(\frac{\tau_{i}}{2}\right).
\]
An analogous decomposition applies to the $\Pi$ and $\Sigma$ matrices:
\begin{align*}
\Pi & =\sum_{j=0}^{3}\Omega_{j}\left(\frac{\tau_{j}}{2}\right)\\
\Sigma & =\sum_{j=0}^{3}\sigma_{j}\left(\frac{\tau_{j}}{2}\right).
\end{align*}
Since $\rho$, $\Pi$ and $\Sigma$ are Hermitian matrices, all their
components in the $\left\{ \tau_{0},\boldsymbol{\tau}\right\} $ basis
are real. Taking into account the commutation and anti-commutation
algebra of Pauli matrices:
\begin{align*}
\left[\tau_{i},\tau_{j}\right] & =2i\epsilon_{ijk}\tau_{k}\,\,\,\,\,\,i,j,k=1,2,3\\
\left\{ \tau_{i},\tau_{j}\right\}  & =2\delta_{ij}I\,\,\,\,\,\,i,j,k=0,\dots,3,
\end{align*}
one can readily prove the evolution equation for the four dimensional
spin-like vector $S=\left(S_{0},\mathbf{S}\right)$ in terms of the
four dimensional vectors $\Omega=\left(\Omega_{0},\mathbf{\Omega}\right)$
and $\sigma=\left(\sigma_{0},\mathbf{\boldsymbol{\sigma}}\right)$
:
\begin{align}
\frac{dS_{0}}{dz} & =-\sigma_{0}S_{0}-\mathbf{\boldsymbol{\sigma}}\cdot\mathbf{S}\nonumber \\
\frac{d\mathbf{S}}{dz} & =-\mathbf{\Omega}\times\mathbf{S}-\sigma_{0}\mathbf{S}-S_{0}\boldsymbol{\mathbf{\sigma}}.\label{eq:spin_equations}
\end{align}
All these vector components can be found out of the elements of their
generating matrices (\ref{eq:pi_sigma_decompositoin}) by using proper
projections. The algebra of $\left\{ \tau_{0},\boldsymbol{\tau}\right\} $
matrices provides the adequate projection operators by means of suitable
tracing: $\Omega_{j}=\mathrm{Tr}\left[\Pi\tau_{j}\right]$ and $\sigma_{j}=\mathrm{Tr}\left[\Sigma\tau_{j}\right]$.
So that:
\begin{align}
\boldsymbol{\Omega} & =\left[\begin{array}{c}
q_{0}+q_{0}^{*}\\
i\left(q_{0}-q_{0}^{*}\right)\\
\mu'_{p}-\mu'_{s}
\end{array}\right]=\left[\begin{array}{c}
\overline{q}'+q'\\
\delta\overline{q}''-\delta q''\\
\mu'_{p}-\mu'_{s}
\end{array}\right]\nonumber \\
\boldsymbol{\sigma} & =\left[\begin{array}{c}
-i\left(\Delta_{0}-\Delta_{0}^{*}\right)\\
\Delta_{0}+\Delta_{0}^{*}\\
l+g
\end{array}\right]=\left[\begin{array}{c}
\delta\overline{q}''+\delta q''\\
-\overline{q}'+q'\\
l+g
\end{array}\right],\label{eq:omega_sigma_vectors}
\end{align}
along with 
\begin{eqnarray*}
\Omega_{0} & = & \mu'_{p}+\mu'_{s}\\
\sigma_{0} & = & l-g.
\end{eqnarray*}

As it is well known, an identical procedure leads to the definition
of the four dimensional real spin vector $S=\left(S_{0},\mathbf{S}\right)$
in terms of the original variational 2D complex vector $C=\left(C_{p},C_{s}\right)^{\top}$:
\begin{equation}
\boldsymbol{S}=\left[\begin{array}{c}
C_{p}^{*}C_{s}+C_{s}^{*}C_{p}\\
i\left(C_{s}^{*}C_{p}-C_{p}^{*}C_{s}\right)\\
\left|C_{p}\right|^{2}-\left|C_{s}\right|^{2}
\end{array}\right]=\left[\begin{array}{c}
2|C_{p}||C_{s}|\cos\phi_{sp}\\
-2|C_{p}||C_{s}|\sin\phi_{sp}\\
\left|C_{p}\right|^{2}-\left|C_{s}\right|^{2}
\end{array}\right],\label{eq:spin_vector}
\end{equation}
where $\phi_{sp}=\phi_{p}-\phi_{s}$ is the relative phase between
the plasmon and soliton components. Additionally, 
\[
S_{0}=\left|C_{p}\right|^{2}+\left|C_{s}\right|^{2}=\left|\mathbf{S}\right|.
\]
Because of the latter constraint, the first evolution equation for
$S_{0}$ in Eqs.(\ref{eq:spin_equations}) is not independent. It
can be obtained from the second one for $\mathbf{S}$ by scalarly
multiplying the latter by $\mathbf{S}$ and by taking into account
the constraint $S_{0}=\left|\mathbf{S}\right|$.

It is interesting to remark here the different nature and physical
meaning of the \textbf{$\Omega$ }and $\sigma$ 4-vectors in the spin
model represented by Eqs.(\ref{eq:spin_equations}). In the absence
of the contribution of the \textbf{$\sigma$ }4-vector, the system
becomes a standard model of a conservative spin system interacting
with an external magnetic field $\boldsymbol{\Omega}$. This model
is a well-kwon representation of a conservative quantum two-level
system, in this case represented by the vector $\left|C\right\rangle =\left(C_{p},C_{s}\right)^{\top}$.
Note, however, that in our case, even in the conservative regime in
which $\sigma_{0}=0$ and $\boldsymbol{\sigma}=0$, both the original
and the equivalent spin models are nonlinear in the sense that the
soliton propagation constant $\mu_{s}$ as well as the couplings $q$
and $\overline{q}$ are functions of the soliton amplitude. Particularly,
the nonlinear character of the couplings confers distinctive features
on the dynamics of the system \cite{Eksioglu2011a}. A nonzero value
of the $\sigma=(\sigma_{0},\boldsymbol{\sigma})$ 4-vector is generated
when the Hamiltonian is non-Hermitian, and this fact can be produced
either because its diagonal terms become complex ($l,g\neq0$) or
because its off-diagonal terms are not conjugate of each other ($q^{*}\neq\overline{q}$).
In this way, the asymmetry in the coupling ($|q|\neq|\overline{q}|$)
appearing in the variational equations (\ref{eq:NL_oscillator_model})
even in the absence of loss and gain \textemdash{} as demonstrated
in Ref.\cite{Ferrando2013b} one expects $\overline{q}\ll q$ in realistic
cases\textemdash{} is also responsible of the breaking of the Hermitian
character of the dynamics. By simple inspection of the expressions
for $(\sigma_{0},\boldsymbol{\sigma})$ in Eqs.(\ref{eq:omega_sigma_vectors})
we recognize that in the absence of gain and loss ($l=g=0$) a zero
value for its zero and third component is obtained ($\sigma_{0}=\sigma_{3}=0$),
whereas the ``symmetry-conjugate'' condition ($q^{*}=\overline{q}$)
ensures the vanishing of its other two components ($\sigma_{1}=\sigma_{2}=0$).
Thus the presence of a non-zero $(\sigma_{0},\boldsymbol{\sigma})$
vector in the spin equations can be interpreted as a signature of
the non-Hermiticity of the Hamiltonian for the variational vector
$C$ caused by two different mechanisms: existence of dissipation
and/or gain (when $l,g\neq0)$ and/or asymmetry in the conjugated
coupling (when $q^{*}\neq\overline{q}$) . 

\section{Stationary spin states \label{sec:Stationary-spin-states}}

In the absence of dissipation and gain, stationary states of the soliplasmon
model (\ref{eq:NL_oscillator_model}) have been found previously for
both symmetric \cite{Bliokh2009,Eksioglu2011a} and asymmetric coupling
\cite{Milian2012a}. These solutions are interpreted as bound states
of a SPP and spatial soliton, the so-called soliplasmons, as they
are characterized by a real propagation constant. This interpretation
is confirmed by numerical simulations of full Maxwell's equations
\cite{Milian2012a} and by a theoretical variational approach \cite{Ferrando2013b}.
Stationary states of the soliplasmon model represent solutions of
full vector nonlinear Maxwell's equations, i.e., they provide a variational
approximation to a monochromatic and stationary 2D nonlinear wave.
In this way, the soliplasmon can be considered, in turn, a soliton,
in which diffraction and nonlinearity subtly cancel each other to
generate a stationary state. 

\begin{figure*}
\hfill{}\includegraphics[width=1\textwidth]{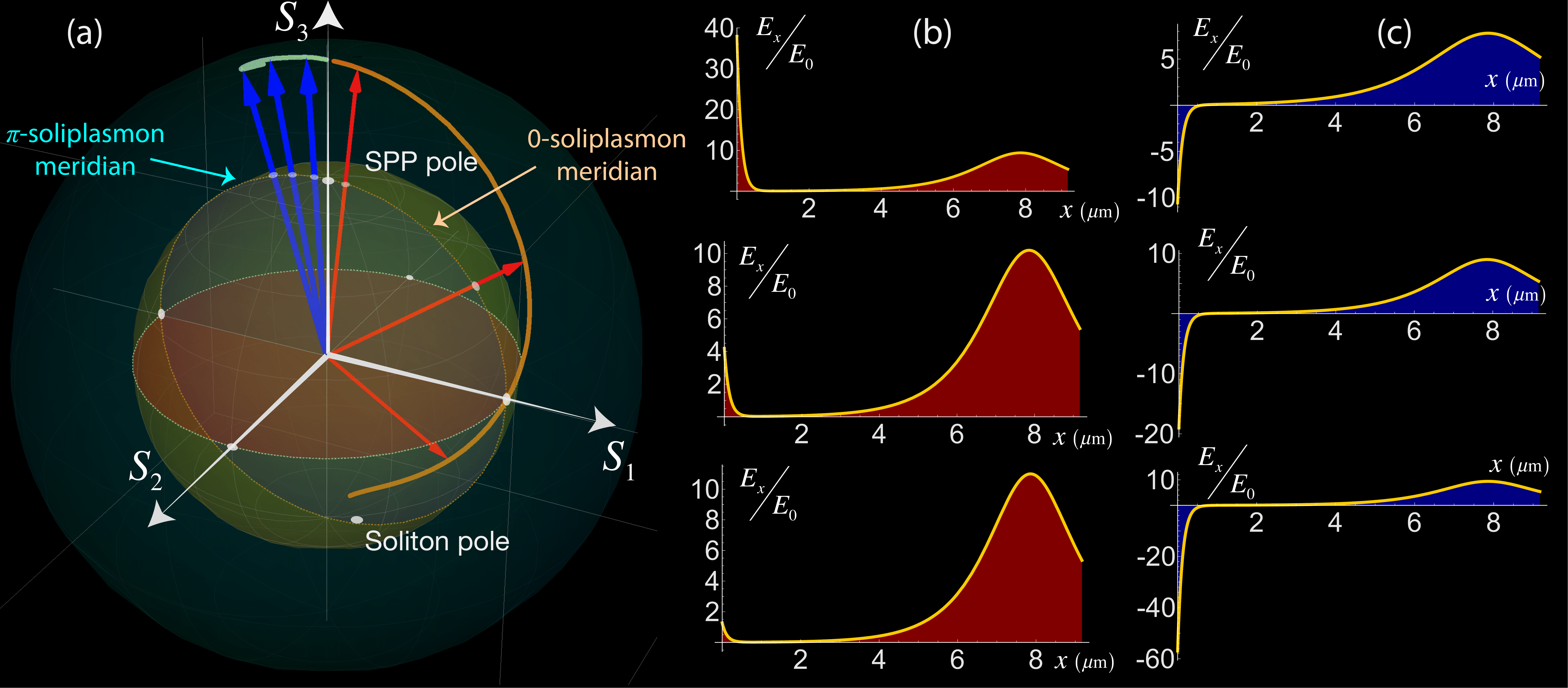}\hfill{}

\caption{Non-dissipative soliplasmons: (a) $0-$ and $\pi-$soliplasmon families
in the Earth-like Poincaré sphere {[}red arrows correspond to representative
$0-$soliplasmons solutions, blue arrows to $\pi-$soliplasmons ones{]};
(b) intensity and phase of the three $0-$soliplasmons solutions (red
arrows) in the Poincaré sphere; (c) the same for the three $\pi-$soliplasmons
solutions (blue arrows). \label{fig:Conserv-soliplasmons-in-Earth-Poincare}}
\end{figure*}

In Figs.~\ref{fig:Conserv-soliplasmons-in-Earth-Poincare}(b) and
\ref{fig:Conserv-soliplasmons-in-Earth-Poincare}(c) we provide some
characteristic examples of the two families of soliplasmon solutions
provided by the variational model (\ref{eq:NL_oscillator_loss_gain})
with the \emph{ansatz} (\ref{eq:variational_ansatz-1}) in the absence
of dissipation and gain. The so-called 0-soliplasmon solutions in
Fig.~\ref{fig:Conserv-soliplasmons-in-Earth-Poincare}(b) constitute
a family characterized by a relative phase $\phi_{sp}=0$ whereas
the $\pi$-soliplasmon family in Fig.~\ref{fig:Conserv-soliplasmons-in-Earth-Poincare}(c)
is characterized by a relative phase $\phi_{sp}=\pi$. According to
the \emph{ansatz }(\ref{eq:variational_ansatz-1})\emph{ }and the
definition of $e_{p}$\emph{ }(\ref{eq:def_ep})\emph{,} the peak
value of the electric field at the metal interface is given approximately
by $C_{p}/\varepsilon_{d}$ whereas the soliton peak value can be
directly approximated by $C_{s}$. Since the variational soliplasmon
model, when applied to a single MDK structure, has no free parameters
we have chosen plausible values for the physical constants of the
MDK structure. We take the Drude model value for the dielectric constant
of silver at $\lambda=765$ nm ($\varepsilon_{m}=-29.95$), $n_{d}=2.2$
for the linear dielectric, and we consider the chalcogenide glass
$\text{As}_{2}\text{Se}_{3}$ as the Kerr medium with $n_{2}=3\times10^{-18}\,\mathrm{m^{2}/W}$
and $n_{K}=2.4$. The choice of $\text{As}_{2}\text{Se}_{3}$ is justified
by the fact that observation of spatial solitons have been recently
reported in slabs of this material with relatively moderate peak intensities
of approximately $I_{0}=2\,\mathrm{GW/cm^{2}}$, which implies a peak
value for the electric field of $E_{0}\sim8\times10^{7}\,\mathrm{V/m}$
\cite{Chauvet2009a} (the electric field in Figs.~\ref{fig:Conserv-soliplasmons-in-Earth-Poincare}(b)
and \ref{fig:Conserv-soliplasmons-in-Earth-Poincare}(c) is normalized
to this value) with an induced nonlinear refractive index of $\Delta n_{\mathrm{NL}}\sim6\times10^{-6}$.
The soliton position is $a=7.3\,\mu\mathrm{m}$ and the width of the
dielectric slab is $d=122\,\mathrm{nm}$. In any case, the purpose
of the variational model is to provide a qualitative approximation
to extant nonlinear solutions rather than an accurate description
of them. It is in this sense that the values taken for the MDK structure
intend to be close to realistic parameters. 

When metal losses are naturally included in the model, soliplasmons
cease to be stationary solutions since their plasmon component automatically
``feels'' the metal losses and it is exponentially attenuated during
propagation. Such as pictorially reflected in Fig. \ref{fig:MDK_energy_flux}(a),
even in the presence of plasmonic losses, a nontrivial \emph{net}
exchange of energy from the soliton to the plasmon component of the
soliplasmon occurs. As full simulations of nonlinear Maxwell's equations
unveil, the pattern of energy exchange crucially depends on phase
initial conditions in such a way attenuation of the plasmon component
can be moderately compensated by energy pumped from the soliton component
\cite{Milian2012a}. This feature is also successfully captured by
the soliplasmon model when $l\neq0$ (see also \cite{Milian2012a}).
If partial compensation of plasmonic losses is achieved by soliton
pumping when no gain is included in the system, a legitimate question
to be considered is if total compensation can be fulfilled if we allow
the system to obtain some gain in the region where the soliton is
localized. In our generalized soliplasmon model this possibility is
naturally incorporated since we precisely assumed the gain to be spatially
localized in the nonlinear region. Therefore, we only need to let
$l\neq0$ and $g\neq0$ in our loss and gain variational equations
(\ref{eq:NL_oscillator_loss_gain}). The question now is whether the
interplay between loss and gain will be able to generate new stationary
solutions for which diffraction and nonlinearity will achieve again
a new perfect balance. 

Since we have proven in Section~\ref{sec:Spin-model} that the loss
and gain soliplasmon model (\ref{eq:NL_oscillator_loss_gain}) is
equivalent to the spin model (\ref{eq:spin_equations}), the question
of the perfect compensation of losses by soliton gain can be reformulated
in terms of finding stationary spin solutions of these spin equations.
The stationarity condition for the solutions of the loss and gain
soliplasmon model (\ref{eq:NL_oscillator_loss_gain}) requires that
both the modulus of the variational parameters $C_{p}$ and $C_{s}$
as well as their relative phase remain $z$-independent. In terms
of its associated spin vector $\boldsymbol{S}$ (\ref{eq:spin_vector})
this condition is equivalent to $d\boldsymbol{S}/dz=0$, which, according
to the spin equation Eq.(\ref{eq:spin_equations}), implies the
\begin{align*}
\boldsymbol{\Omega}\times\boldsymbol{S}+\sigma_{0}\boldsymbol{S}+S_{0}\boldsymbol{\sigma} & =0
\end{align*}
or
\begin{equation}
\boldsymbol{\Omega}\times\boldsymbol{n}+\sigma_{0}\boldsymbol{n}+\boldsymbol{\sigma}=0,\label{eq:stationary_eqs_spin}
\end{equation}
where we have introduced the unitary vector $\boldsymbol{n}\equiv\boldsymbol{S}/|\boldsymbol{S}|=\boldsymbol{S}/S_{0}$.

Remarkably, the previous equation for $\boldsymbol{n}$ admits the
following explicit solution:
\begin{equation}
\boldsymbol{n}=-\frac{1}{\sigma_{0}}\left(\frac{1}{\sigma_{0}^{2}+|\boldsymbol{\Omega}|^{2}}\right)\left[\sigma_{0}^{2}\boldsymbol{\sigma}+\left(\boldsymbol{\sigma}\cdot\boldsymbol{\Omega}\right)\boldsymbol{\Omega}+\sigma_{0}\left(\boldsymbol{\sigma}\times\boldsymbol{\Omega}\right)\right].\label{eq:stationary_sol_n}
\end{equation}
Inasmuch as $\boldsymbol{n}$ is unitary, the unitarity condition
$\boldsymbol{n}\cdot\boldsymbol{n}=1$ introduce the following important
constraint on the $(\sigma_{0},\boldsymbol{\sigma})$ and $(\Omega_{0},\boldsymbol{\Omega})$
parameters:
\begin{equation}
\sigma_{0}^{2}|\boldsymbol{\sigma}|^{2}+\left(\boldsymbol{\sigma}\cdot\boldsymbol{\Omega}\right)^{2}=\sigma_{0}^{2}\left(\sigma_{0}^{2}+|\boldsymbol{\Omega}|^{2}\right).\label{eq:constraint_sigma_omega}
\end{equation}
 This constraint is a necessary condition to be fulfilled by the parameters
of the spin model for the existence of the solution (\ref{eq:stationary_sol_n}).
It is a highly remarkable feature that this constraint is absent in
the conservative case for which $\sigma_{0}=0$ and $\boldsymbol{\sigma}=0$. 

In order to illustrate the relevance of the constraint (\ref{eq:constraint_sigma_omega})
it is instructive to approach the conservative regime by taking the
limit $\sigma=(\sigma_{0},\boldsymbol{\sigma})\rightarrow0$ in Eq.(\ref{eq:stationary_sol_n})
and (\ref{eq:constraint_sigma_omega}). We easily find that:
\begin{align*}
\boldsymbol{n} & =-\frac{\left(\boldsymbol{\sigma}\cdot\boldsymbol{\Omega}\right)\boldsymbol{\Omega}}{\sigma_{0}|\boldsymbol{\Omega}|^{2}}+O(\sigma)\text{\,\,\,\ and}\\
\left(\boldsymbol{\sigma}\cdot\boldsymbol{\Omega}\right) & =\pm\sigma_{0}|\boldsymbol{\Omega}|+O(\sigma)^{3},
\end{align*}
respectively. So that, by incorporating the approximated constraint
in the equation for $\boldsymbol{n}$, we get:
\begin{equation}
\boldsymbol{n}=\mp\frac{\boldsymbol{\Omega}}{|\boldsymbol{\Omega}|}+O(\sigma).\label{eq:conservative_solution}
\end{equation}
This is the correct solution in the conservative case, in which the
stationary solution correspond to a spin perfectly aligned with the
external magnetic field $\boldsymbol{\Omega}$. As we can see, when
taking the conservative limit the correct result is obtained only
when one properly complements the solution with the constraint (\ref{eq:constraint_sigma_omega}).
On the other hand, as already emphasized, the conservative solution
(\ref{eq:conservative_solution}) \textemdash unlike the non-conservative
spin solution (\ref{eq:stationary_sol_n})\textemdash{} automatically
satisfies the unitarity condition $\boldsymbol{n}\cdot\boldsymbol{n}=1$
independently of the value of the parameters of the spin model.

Another interesting limit is the one corresponding to non-dissipative
solutions ($l=g=0$) in the presence of asymmetric coupling ($q^{*}\neq\overline{q}$)
implying the non-hermiticity of the Hamiltonian. If we neglect metal
losses, this situation corresponds to the more common case of conservative
soliplasmons for which the soliton-to-plasmon coupling is considerably
larger than the plasmon-to-soliton one, and, therefore, coupling is
clearly asymmetric \cite{Milian2012a,Ferrando2013b}. According to
Eqs.(\ref{eq:omega_sigma_vectors}), in this situation we wish to
take the $\sigma_{0}\rightarrow0$ limit \textemdash keeping $\boldsymbol{\sigma}$
finite\textemdash{} in the general solution (\ref{eq:stationary_sol_n})
and in the constraint (\ref{eq:constraint_sigma_omega}). A similar
analysis to the previous one for the full-conservative case leads
us to the following expression for the unitary spin vector $\boldsymbol{n}$
of non-dissipative asymmetrically coupled solutions: 
\begin{align}
\boldsymbol{n} & =\frac{1}{\left|\boldsymbol{\Omega}\right|^{2}}\left[\mp\left(\left|\boldsymbol{\Omega}\right|^{2}-\left|\boldsymbol{\sigma}\right|^{2}\right)^{1/2}\boldsymbol{\Omega}-\boldsymbol{\sigma}\times\boldsymbol{\Omega}\right]\label{eq:assymetric_coupling_solution_n}
\end{align}
together with the constraint for the spin model parameters $\boldsymbol{\sigma}\cdot\boldsymbol{\Omega}=0$.
Note that the previous expression reduces to the conservative one
(\ref{eq:conservative_solution}) when we take the hermiticity limit
$\boldsymbol{\sigma}\rightarrow0$. 

\section{Stationary dissipative soliplasmons and the ``golden constraint''\label{sec:Stationary-dissipative-soliplasm}}

The previous analysis has no explicit reference to he dependence of
the vectors $(\sigma_{0},\boldsymbol{\sigma})$ and $\boldsymbol{\Omega}$
on the parameters or our original dissipative soliplasmon model (\ref{eq:NL_oscillator_loss_gain}).
In this sense, its validity goes beyond the specificities of our model
describing soliplasmon dynamics. However, as we will see next, the
specific properties of the dissipative soliplasmon model are important.
The set of equations (\ref{eq:omega_sigma_vectors}) provide the link
between the generic spin model parameters and the matrix elements
of the soliplasmon Hamiltonian (\ref{eq:hamiltonian}). In order to
understand the peculiar properties of the spin model of dissipative
soliplasmons one has to keep in mind that all the coefficients of
the Hamiltonian (\ref{eq:hamiltonian}) except $\mu_{p}$ are nonlinear
functions of the $z$-dependent soliton variational parameter $C_{s}(z)$.

\begin{figure*}
\includegraphics[width=0.8\textwidth]{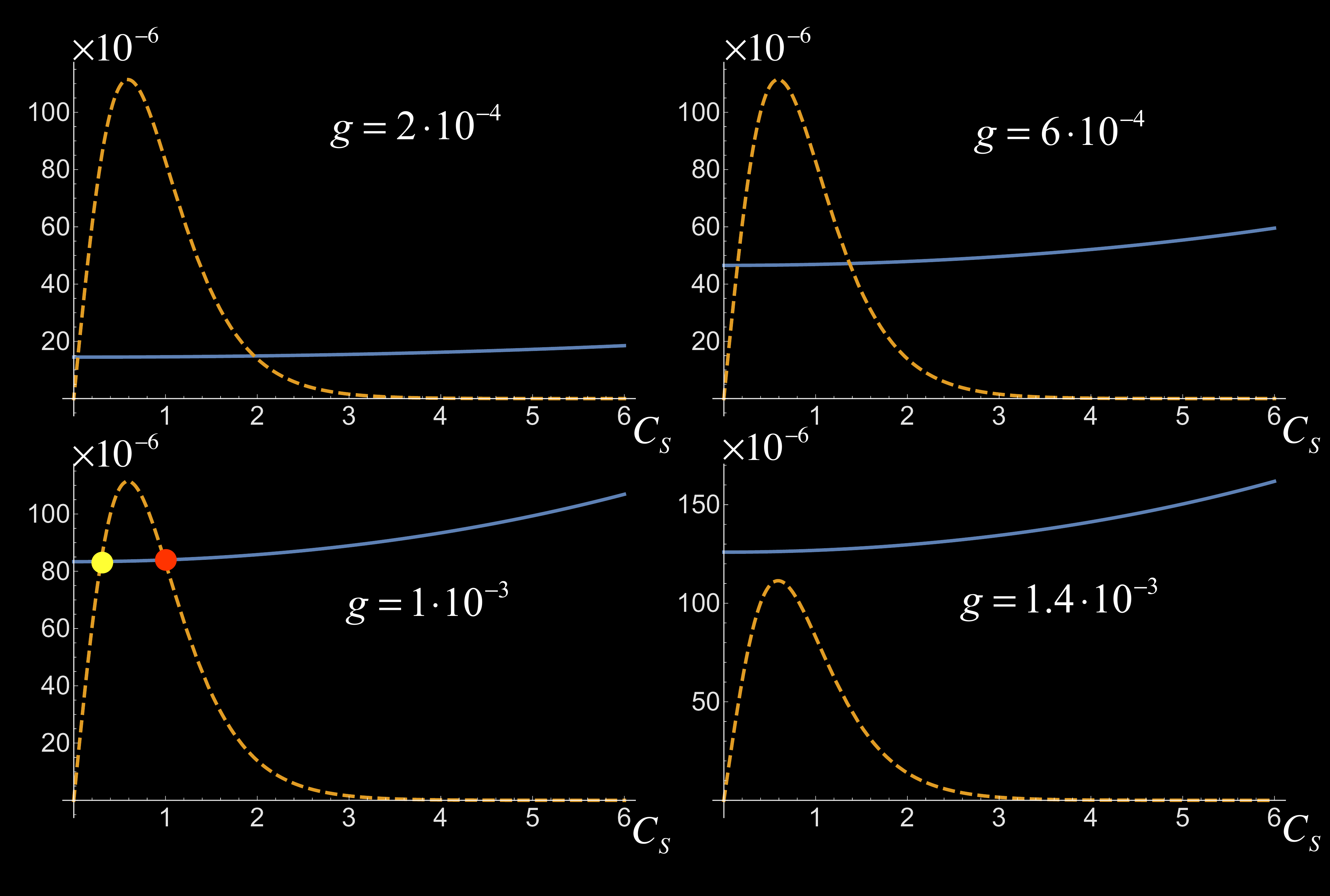}

\caption{Representation of the ``golden constraint'' for increasing values
of the gain coefficient $g$ for a given loss coefficient of $l=1.05\times10^{-2}$
($l=758\,\mathrm{cm}^{-1}$, in physical units).\label{fig:golden_constraint} }
\end{figure*}

\subsection{The ``golden constraint'' for nonlinear solutions}

We have stressed already in Section~\ref{sec:Stationary-spin-states}
the importance of the constraint (\ref{eq:constraint_sigma_omega})
to determine a legitimate spin solution. In terms of our original
soliplasmon model this constraint becomes a condition to be fulfilled
also by the parameters of the soliplasmon Hamiltonian (\ref{eq:hamiltonian}).
By substituting the values of the $(\sigma_{0},\boldsymbol{\sigma})$
and $(\Omega_{0},\boldsymbol{\Omega})$ vectors in the constraint
(\ref{eq:constraint_sigma_omega}) using Eqs.(\ref{eq:omega_sigma_vectors})
we obtain that the coefficients of the dissipative soliplasmon model
(\ref{eq:NL_oscillator_loss_gain}) have to fulfill the following
constraint: 
\begin{multline}
\left[(g+l)(\mu'_{p}-\mu'_{s})-(\delta\overline{q}''-\delta q'')(\overline{q}'-q')\right]^{2}-\\
(g-l)^{2}\left[-4gl+(\mu'_{p}-\mu'_{s})^{2}+\right.\\
\left.4q'\overline{q}'+(\delta\overline{q}''-\delta q'')^{2}\right]=0.\\
\label{eq:golden_constraint_general}
\end{multline}
 Since we are dealing with the imaginary part $\varepsilon_{L}''(x)$
as a small perturbation, the imaginary parts of the couplings $\delta q''$
and $\delta\overline{q}''$ have to be considered, in turn, small
perturbation of their real parts $q'$ and $\overline{q}'$. On the
other hand, the variational equations of the soliplasmon model were
derived under the weak coupling condition, so the coupling coefficients
$q$ and $\overline{q}$ were assumed to be small with respect the
propagation constants $\mu_{p}$ and $\mu_{s}$. Consequently, we
can further simplify Eq.(\ref{eq:golden_constraint_general}) by keeping
only the leading order terms in both approximations, which in this
context implies neglecting the imaginary parts of the couplings $\delta q''$
and $\delta\overline{q}''$ . In this way we obtain a simple form
of the constraint given by:
\begin{multline}
gl\left\{ \left(g-l\right)^{2}+\left[\mu'_{p}-\mu'_{s}\left(C_{s}\right)\right]^{2}\right\} -\\
\left(g-l\right)^{2}q'(C_{s})\overline{q}'(C_{s})=0,\label{eq:golden_constraint}
\end{multline}
where we have introduced the dependence on the soliton variational
parameter $C_{s}$ explicitly.

Written in the form (\ref{eq:golden_constraint}) it becomes clear
the essential role played by the constraint equation, either in its
simplified or general form, in the determination of the nonlinear
solutions of a dissipative soliplasmon. Indeed, when all the non-variational
parameters of the problem are given, the constraint equation define
whether a solution can exist or not. The constraint is nothing but
a nonlinear equation for the modulus of the soliton parameter $|C_{s}|$.
Specifically for the MDK structure under consideration, if we provide
as an input its optical complex coefficients $\varepsilon_{m}$, $\varepsilon_{d}$,
$\varepsilon_{K}$, the nonlinear index $n_{2}$, the width $d$,
and the soliton position $a$, we can compute all terms in the constraint
using the expressions given in the Appendix A \textemdash including
the loss and gain coefficients $l$ and $g$\textemdash{} except the
unknown value of $|C_{s}|$. Since this constraint is a necessary
condition for a stationary spin solution as given in Eq.(\ref{eq:stationary_sol_n})
to exist, if such a solution for $|C_{s}|$ cannot be found, we can
state that the corresponding dissipative soliplasmon solution does
not exist either. On the contrary, if one solution $|C_{s}|$ of the
constraint is found we can univocally construct its associated spin
$\boldsymbol{S}$ and variational $C$ vectors. Once the value of
$|C_{s}|$ is fixed by the constraint in terms of the system parameters,
all the nonlinear coefficients $\mu'_{s}$, $q$ and $\overline{q}$
depending on it are likewise fixed. It is interesting to emphasize
that, due to the nonlinear character of the constraint (\ref{eq:golden_constraint}),
it is possible to find more than one solution for $|C_{s}|$.

Due to the paramount importance of the nonlinear constraint (\ref{eq:golden_constraint})
\textemdash or (\ref{eq:golden_constraint_general})\textemdash{}
for finding dissipative soliplasmon solutions we shall refer to it
as the ``golden constraint''. Remarkably, the ``golden constraint''
only exists in the presence of dissipation. When we set $l=g=0$ in
it, the constraint disappears and, therefore, no restriction is exerted
on the nonlinear solutions of the problem. In the same way that the
existence of conservative spatial solitons is rooted in the compensation
between diffraction and nonlinearity, the ``golden constraint''
can be envisaged as the additional balance condition between losses
and gain that a dissipative soliton has to fulfill in order to exist
\cite{Akhmediev2008a}. Unlike in the conservative case, in which
soliton solutions constitute a continuos family, the ``golden constraint''
permit only a discrete number of them.

\subsection{Spin solution of a stationary dissipative soliplasmon}

We are now in conditions to find an explicit expression for a spin
solution of a stationary dissipative soliplasmon. Firstly, we substitute
in the general stationary spin solution (\ref{eq:stationary_sol_n})
the explicit expressions of the $\left(\sigma_{0},\boldsymbol{\sigma}\right)$
and $\left(\Omega_{0},\boldsymbol{\Omega}\right)$ vectors in terms
of the coefficients of the dissipative soliplasmon model as given
in Eqs.(\ref{eq:omega_sigma_vectors}). Secondly, we implement the
``golden constraint'' (\ref{eq:golden_constraint}) explicitly in
the solution by eliminating the dependence on the propagation constant
``detuning'' $\mu'_{p}-\mu'_{s}$ in favor of a dependence on $l$,
$g$, $q$ and $\overline{q}$. The result is:
\begin{align}
\boldsymbol{n} & =\left[\begin{array}{c}
2\mathrm{sgn}\left(l-g\right)\left(gq+l\overline{q}\right)^{-1}\left[\left(q\overline{q}-gl\right)gl\right]^{1/2}\\
\\
-2gl\left(gq+l\overline{q}\right)^{-1}\\
\\
-1+2gq\left(gq+l\overline{q}\right)^{-1}
\end{array}\right].\label{eq:explicit_sol_n}
\end{align}
We stress again that the problem of finding a dissipative soliplasmon
solution is univocally solved once we have determined the specific
value of $|C_{s}|$ that satisfies the ``golden constraint''. This
can be seen clearly in the previous expression, in which once $|C_{s}|$
is given $q$ and $\overline{q}$ are determined and the spin solution
is fixed. The particular form of (\ref{eq:explicit_sol_n}) is, however,
not unique since we can implement de ``golden constraint'' differently
from what we did to obtain this solution. We could have chosen to
eliminate other parameter rather than the detuning. In that case the
explicit form of $\boldsymbol{n}$ in Eq.(\ref{eq:explicit_sol_n})
would look different although both expressions would correspond to
the same solution. 

The spin solution $\boldsymbol{n}$ in Eq.(\ref{eq:explicit_sol_n})
is a unit vector, for which $S_{0}=|\boldsymbol{n}|^{2}=1$. Therefore,
it does not contain information about the norm of the variational
vector $C$, which is given by the zero component of the spin vector
$S_{0}=|C_{p}|^{2}+|C_{s}|^{2}$. In order to define the spin solution
of a dissipative soliplasmon completely we need to give an explicit
expression of $S_{0}$ in terms of known parameters. From the definitions
of $S_{0}$ and $S_{3}$ in Eq.(\ref{eq:spin_vector}), it is immediate
to obtain:
\begin{equation}
S_{0}=\frac{2|C_{s}|^{2}}{1-n_{3}}.\label{eq:S0}
\end{equation}
This expression together with the unit vector (\ref{eq:explicit_sol_n})
define completely the spin solution of a dissipative soliplasmon since
$\boldsymbol{S}=S_{0}\boldsymbol{n}.$

\begin{figure*}
\includegraphics[width=0.9\textwidth]{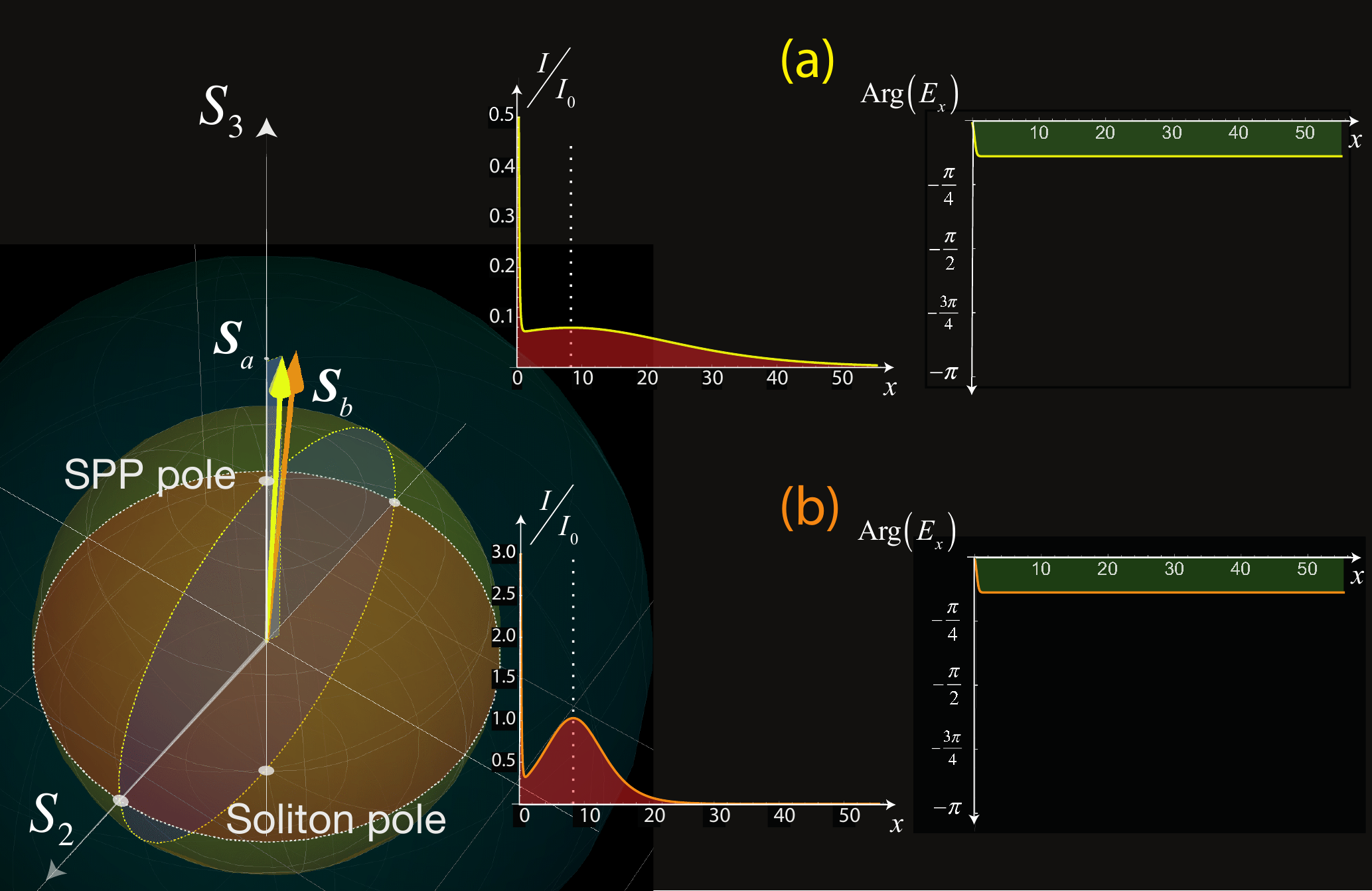}

\caption{The pair of dissipative soliplasmons in the ``Earth-like'' Poincaré
sphere corresponding to the configuration with $g=1\times10^{-3}$
($g=72\,\mathrm{cm}^{-1}$, in physical units) in Fig.~\ref{fig:golden_constraint}:
$\boldsymbol{S}_{a}$ (yellow arrow) solution with $C_{s}=0.295$;
$\boldsymbol{S}_{b}$ (orange arrow) solution with $C_{s}=0.848$.
\emph{Inset}: Intensity and phase profiles of the corresponding two
dissipative soliplasmon solutions represented in the ``Earth-like''
Poincaré sphere: (a) $\boldsymbol{S}_{a}$ yellow arrow solution;
(b) $\boldsymbol{S}_{b}$ orange arrow solution.\label{fig:Intensity-and-phase_dissipative_soliplamons}}
\end{figure*}

The Poincaré or Bloch sphere is a common way of representing the spin
vector associated to a two-level quantum system. We will use it to
represent soliplasmon states given by the previous spin equations.
Inasmuch as Eq.(\ref{eq:S0}) points out that soliplasmon states represented
by spin vectors with arbitrarily large norm $S_{0}$ can exist (if
$n_{3}\rightarrow1$), we will extend the Poincaré sphere in order
to guarantee the visualization of these states. We introduce the ``Earth-like''
Poincaré sphere concept, which maps the infinite existence domain
of $S_{0}$, given by $\left[0,\infty\right[$, into a finite one,
$\left[0,\varphi\right]$ where $\varphi\approx1.618$ is the \emph{golden
ratio}, to facilitate this visualization. Details of this construction
are given in Appendix B. By definition, the ``Earth-like'' Poincaré
sphere is defined by two regions: the inner sphere (the Earth's sphere),
characterized by the transformed radius $\bar{S}_{0}\in\left[0,1\right]$,
and the outer shell given by the domain $\bar{S}_{0}\in\left]1,\varphi\right[$
(the ``atmospheric'' layer.) All spins whose end point lies on the
``atmosphere'' layer will have $S_{0}>1$. The main advantage of
this representation is that existing states with infinitely large
norm approaching the $S_{0}\rightarrow\infty$ limit will appear represented
by points infinitely close to the atmosphere's outer surface with
radius $\varphi$ (i.e., $\bar{S}_{0}\rightarrow\varphi$). 

As a first example of its use, we provide in Fig.~\ref{fig:Conserv-soliplasmons-in-Earth-Poincare}(a)
the spin representation of the non-dissipative $0-$soliplasmons (red
arrows) and $\pi$-soliplasmons (blue arrows) of Figs.~\ref{fig:Conserv-soliplasmons-in-Earth-Poincare}(b)
and \ref{fig:Conserv-soliplasmons-in-Earth-Poincare}(c). The spin
vectors for these solutions are given by Eqs.(\ref{eq:assymetric_coupling_solution_n})
and (\ref{eq:S0}). Non dissipative soliplasmons are restricted to
lye on the $S_{2}=0$ section of the sphere since its relative phase
is either $0$ or $\pi$. The $0-$ and $\pi-$ soliplasmon families
form a continuum of solutions, which are represented in this figure
by the orange and light blue curves, respectively. However, the norm
of these solutions change with $C_{s}$, as clearly visualized in
Fig.~\ref{fig:Conserv-soliplasmons-in-Earth-Poincare}(a). The ``Earth
surface'' has unity radius and it has been selected in order that
soliplasmons with $S_{3}=0$, i.e., with $C_{p}=C_{s}$, lie on it.
The north pole of the unit sphere is given by the \emph{plasmonic}
state $C_{p}=1$ and $C_{s}=0$. However, both the $0-$ and $\pi-$soliplasmon
curves tend to the north pole of the \emph{outer} sphere indicating
that for non-dissipative soliplasmons these asymptotic states have
infinitely large plasmonic component ($C_{p}\rightarrow\infty$).
An analogous interpretation could be given to the southern pole with
increasingly solitonic states.

\section{Critical gain\label{sec:Stationary-critical-gain}}

We will use the ``Earth-like'' Poincaré sphere also to represent
dissipative soliplasmons given by the spin vector defined by Eqs.(\ref{eq:explicit_sol_n})
and \ref{eq:S0}. The main difference with respect to the non-dissipative
case is certainly the existence of the ``golden constraint'' (\ref{eq:golden_constraint})
(we will focus on the simple version), which restricts the number
of potential solutions to a discrete set instead of a continuous family.
The number of solutions is determined by the finite number of roots
in the $C_{s}$ variable given by Eq. (\ref{eq:golden_constraint}).
In order to have an idea of this number, it is convenient to rewrite
(\ref{eq:golden_constraint}) in the form:
\begin{equation}
q'(C_{s})\overline{q}'(C_{s})=gl\left\{ 1+\left[\frac{\mu'_{p}-\mu'_{s}\left(C_{s}\right)}{g-l}\right]^{2}\right\} .\label{eq:golden_constraint_rewritten}
\end{equation}

It can be checked that the product $q'\overline{q}'$ behaves approximately
as $k_{1}\left|C_{s}\right|\exp(-k_{2}\left|C_{s}\right|)$, so that
it presents the characteristic shape shown by the dashed curve in
Fig.~\ref{fig:golden_constraint}. In this representation we use
the same parameters for the MDK structure as in the conservative case
but taking now a realistic value for the complex silver dielectric
constant ($\varepsilon_{m}=-34.0+1.8i$ at $\lambda=870\,\mathrm{nm}$),
which determines a SPP loss dimensionless coefficient of $l=1.05\times10^{-2}$
(corresponding to $l=758\,\mathrm{cm}^{-1}$, in physical units.)
The function at the right hand side of Eq.(\ref{eq:golden_constraint_rewritten})
is a slowly monotonically increasing function of $C_{s}$ parametrized
by the gain coefficient $g$ (solid line in Fig.~\ref{fig:golden_constraint}.)
The crossing points are the roots of the ``golden constraint'' fixing
the allowed values of $C_{s}$ for the permitted dissipative soliplasmon
solutions. The smooth gradient of the right hand side function in
Eq.(\ref{eq:golden_constraint_rewritten}) can be considered a general
feature caused by the soft modulation of the induced nonlinear refractive
index with the soliton amplitude for plausible intensities in realistic
nonlinear materials, due to the small value of the nonlinear coefficient
$\gamma$. The sign of the slope depends on the relative sign between
the SPP and soliton paraxial propagation constants. 

The nonlinear character of the ``golden constraint'' determines
the existence of a critical value for the gain. Due to the peculiar
nonlinear dependence of $q'\overline{q}'$ on $\left|C_{s}\right|$,
as compared to the smoothness of the dependence on $\left|C_{s}\right|$
of soliton propagation constant $\mu'_{s}$, the typical scenario
for crossing points is the one depicted in Fig. \ref{fig:golden_constraint}.
As we increase $g$, the monotonically increasing right-hand side
of Eq.(\ref{eq:golden_constraint_rewritten}) (solid line in Fig.\pageref{fig:golden_constraint})
takes a larger value at the origin. Thus, two crossing points are
granted to exist until the solid line intersects the $q'\overline{q}'$
function at its maximum, something that occurs at a critical value
of $g=g_{c}$. Consequently, for values of $g$ below $g_{c}$ there
exist two stationary dissipative soliplasmon solutions. As $g$ tends
to $g_{c}$ the two solutions become increasingly more similar in
such a way they become degenerate at $g=g_{c}$. The solution at $g=g_{c}$
is unique, and no stationary dissipative soliplasmon solution exists
for $g>g_{c}$. We will refer to $g_{c}$ as the\emph{ critical gain}.

In the left-bottom side of Fig.~\ref{fig:Intensity-and-phase_dissipative_soliplamons}
we represent in the ``Earth-like'' Poincaré sphere the two spins,
denoted as $\boldsymbol{S}_{a}$ and $\boldsymbol{S}_{b}$, associated
to the pair of soliplasmon solutions obtained from the roots of the
``golden constraint'' with gain coefficient $g=1\times10^{-3}$
($g=72\,\mathrm{cm}^{-1}$, in physical units) in Fig.~\ref{fig:golden_constraint}.
In this case, the ``Earth surface'' with unity radius has been chosen
in such a way the purely solitonic soliplasmon, i.e., the state with
$C_{p}=0$ and $C_{s}=E_{0}$, lies on it. Thus, by construction,
the spin state $(0,0,-E_{0}^{2})$ is the state whose norm sets the
reference for all the states and it will represented by $\boldsymbol{S}=\left(0,0,-1\right)$
in the ``Earth-like'' Poincaré sphere. The reason for taking this
state as a reference will become clear when we analyze propagation
in the next section. In general, $E_{0}$ is an arbitrary reference
value. Here, as in previous section and for the sake of comparison
with realistic situations, we choose it to be $E_{0}=8\times10^{7}\,\mathrm{V/m}$,
which approximately corresponds to the soliton peak value for realistic
experiments in chalcogenide materials \cite{Chauvet2009a}. 

Dissipative soliplasmons qualitatively differ from the soliplasmon
of the non-dissipative system ($l=g=0$) not only because they constitute
a discrete set of two solutions instead of a family \textemdash compare
Fig.~\ref{fig:Conserv-soliplasmons-in-Earth-Poincare} and the $\boldsymbol{S}_{a}$
and $\boldsymbol{S}_{b}$ solutions in Fig.~\ref{fig:Intensity-and-phase_dissipative_soliplamons}(a)
and (b)\textemdash{} but also because they are not restricted to lie
in the $S_{2}=0$ section of the ``Earth-like'' Poincaré sphere.
Therefore, they are not either $0-$ nor $\pi-$ soliplasmons, as
clearly visualized in Fig.~\ref{fig:Intensity-and-phase_dissipative_soliplamons}
by the two spins $\boldsymbol{S}_{a}$ and $\boldsymbol{S}_{b}$.
The loss and gain balance expressed in the ``golden constraint''
requires the stationary soliplasmon solutions to present a nontrivial
relative phase to achieve perfect compensation. In our particular
case, both solutions present a very similar relative phase close to
$-\pi/8$, as visualized in the ``Earth-like'' Poincaré sphere of
Fig.~\ref{fig:Intensity-and-phase_dissipative_soliplamons}, and
thus an almost identical phase profile, as shown in Fig.~\ref{fig:Intensity-and-phase_dissipative_soliplamons}.
In our particular case, both spin solutions are very close to the
SPP pole and to each other thus indicating they have large SPP components
and similar shapes. The fact that dissipative soliplasmons with large
SPP component exist is important for plasmonic nonlinear amplification,
as we will see in the next section.

\begin{figure*}
\includegraphics[width=0.6\textwidth]{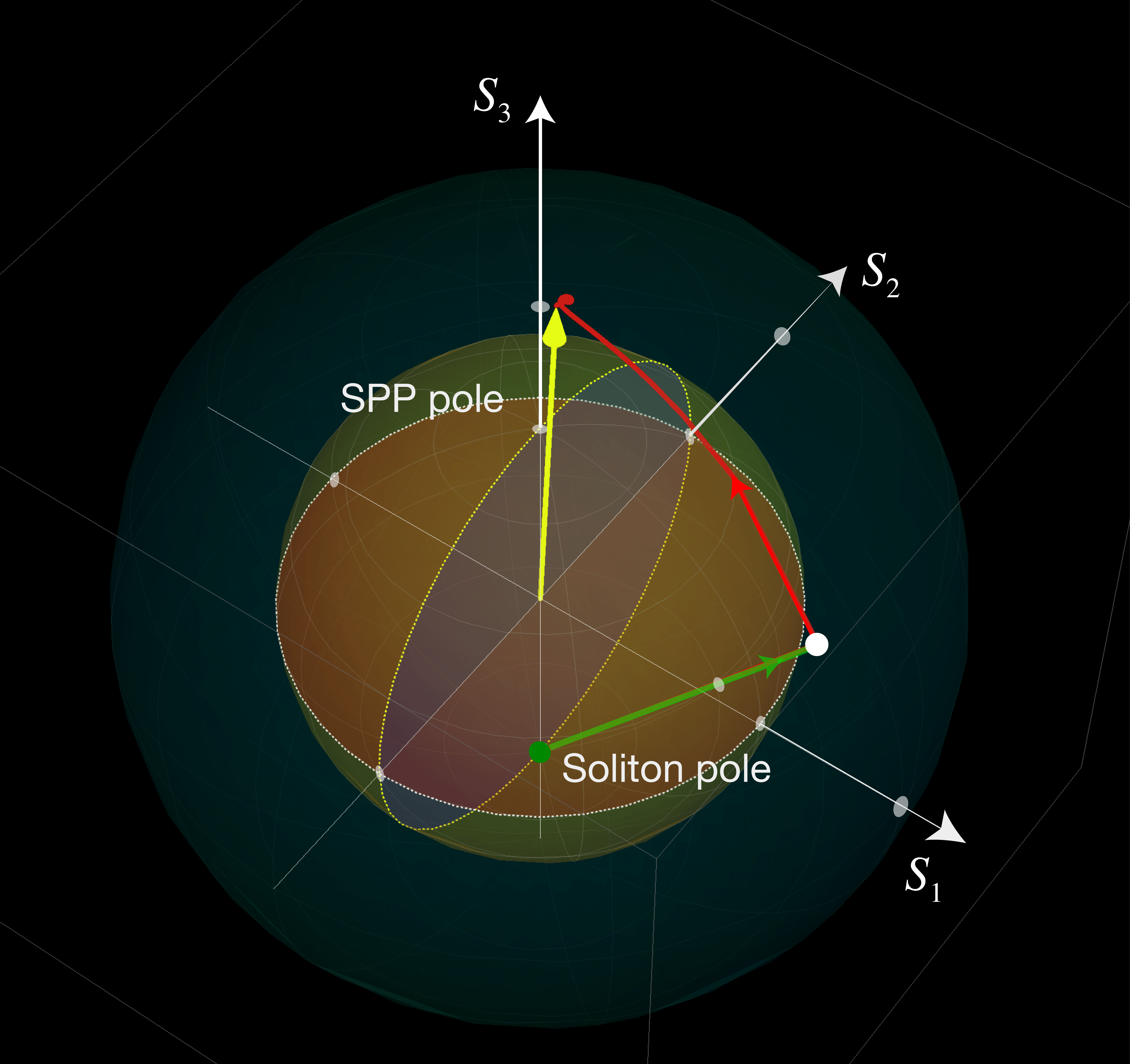}

\caption{Nonlinear amplification of a surface plasmon in the ``Earth-like''
Poincaré sphere. A purely solitonic input state (green dot) with no
plasmonic component at the South pole evolves into a dissipative soliplasmon
output state (yellow arrow) with a high plasmon component located
near the north pole well above the Earth surface. Trajectory ``takes
off'' from the Soliton pole of the Earth surface into the ``atmospheric''
shell swiftly reaching a point of maximum height/amplification (white
spot), then descending smoothly to reach the final output dissipative
soliplasmon state. \label{fig:Nonlinear-amplification_sphere}}
\end{figure*}

The stability mechanism behind these dissipative soliplasmon solutions
is, in fact, very robust, as our simulation in Fig.~\ref{fig:Nonlinear-amplification_sphere}
unveils. Our choice of the initial condition to demonstrate plasmonic
amplification during propagation can be also considered as an stability
check. Our initial purely solitonic state $\boldsymbol{S}^{\mathrm{in}}=\left(0,0,-1\right)$
can be taken as a perturbation of the final soliplasmon state $\boldsymbol{S}_{a}$
(yellow arrow) by writing $\boldsymbol{S}^{\mathrm{in}}=\boldsymbol{S}_{a}+\Delta\boldsymbol{S}_{a}$.
It is clear in the Earth-like Poincaré sphere that, since $\boldsymbol{S}_{a}$
is a ``highly-plasmonic'' state \textemdash very close to the north
pole\textemdash , the perturbation $\Delta\boldsymbol{S}_{a}$ is
huge since the initial state $\boldsymbol{S}^{\mathrm{in}}$ is located
at the south pole. Despite this fact, the dissipative soliplasmon
solution $\boldsymbol{S}_{a}$ acts as an attractor for $\boldsymbol{S}^{\mathrm{in}}$
converting an state with no plasmonic component at all into a state
with a large relative plasmonic component. On the hand, the intermediate
state of maximum amplification (white spot in Fig.~\ref{fig:Nonlinear-amplification_sphere})
is not a stable solution, it is also belongs to the attractor basin
of the stationary dissipative soliplasmon $\boldsymbol{S}_{a}$.

\section{Nonlinear amplification\label{sec:Nonlinear-amplification} }

The existence of dissipative soliplasmons in a MDKD structure with
gain in the Kerr medium provides an original mechanism for the amplification
of surface plasmon polaritons. If a dissipative soliplasmon is the
asymptotic stationary state of an initial state with smaller plasmon
component then amplification of the SPP signal occurs. The soliplasmon
model and its spin representation provides an useful scenario to describe
this mechanism. Using the soliplasmon \emph{ansatz }(\ref{eq:variational_ansatz-1}),\emph{
}we can characterize the SPP amplification property in simple terms
as 
\begin{equation}
\left|C_{p}^{\mathrm{in}}\right|<\left|C_{p}^{\mathrm{out}}\right|.\label{eq:amplification_condition}
\end{equation}
In our spin model, the output state with plasmon component $C_{p}^{\mathrm{out}}$
will be represented by a dissipative soliplasmon spin $\boldsymbol{S}^{\mathrm{out}}$in
the ``Earth-like'' Poincaré sphere. At the same time, the states
verifying the condition (\ref{eq:amplification_condition}) will constitute
in the ``Earth-like'' Poincaré sphere the domain $\boldsymbol{\Omega}_{\mathrm{amp}}\left(\boldsymbol{S}^{\mathrm{out}}\right)$
of initial states that can be \emph{potentially} amplified into $\boldsymbol{S}^{\mathrm{out}}$.
Mathematically, if a solution $\boldsymbol{S}\left(z\right)$ has
$\boldsymbol{S}^{\mathrm{out}}$ as its asymptotic state and, \emph{simultaneously},
its input state $\boldsymbol{S}^{\mathrm{in}}$ belongs to $\boldsymbol{\Omega}_{\mathrm{amp}}\left(\boldsymbol{S}^{\mathrm{out}}\right)$
then it is guaranteed that a nonlinear amplification of the SPP occurs. 

The previous mechanism is clearly visualized in Fig.~\ref{fig:Nonlinear-amplification_sphere}.
In this figure we have represented a paradigmatic example of nonlinear
SPP amplification. Our initial state is a soliplasmon state with no
plasmon component at all located right at the south pole of the Earth
surface, i.e., with $C_{p}^{\mathrm{in}}=0$ and $C_{s}^{\mathrm{in}}=E_{0}$,
where $E_{0}$ is our normalization value chosen as in Sections \ref{sec:Stationary-spin-states}
and \ref{sec:Stationary-dissipative-soliplasm}. Our physical parameters
are the same used to find the dissipative soliplasmon pair in Fig.~\ref{fig:Intensity-and-phase_dissipative_soliplamons}
in previous section. This means that the gain $g$ is kept below the
critical gain $g_{c}$. By solving the evolution spin equations with
this initial condition we find that indeed there is a trajectory that
connects the initial spin state $\boldsymbol{S}^{\mathrm{in}}=\left(0,0,-1\right)$
with the dissipative soliplasmon solution $\boldsymbol{S}_{a}$ (yellow
arrow) represented in Fig.~\ref{fig:Intensity-and-phase_dissipative_soliplamons}.
The evolution process presents two different well defined phases.
First, there is a process of a strong and fast amplification, which
takes the initial spin towards a point of maximal amplification, represented
in Fig.~\ref{fig:Nonlinear-amplification_sphere} by a white spot.
Amplification is visualized in the fact that this point is located
in the atmospheric shell indicating that the norm is higher than unity
and thus larger than the initial one. Besides, this amplification
generates plasmonic component \textemdash absent in the initial state\textemdash{}
since we moved away from the soliton pole towards the northern hemisphere.
The fact that the white spot appears visually close to the outer surface
of the atmospheric shell indicates that amplification is large. The
norm of the white spot is in fact larger than in any other evolution
state, including the final state, thus indicating that is the point
of maximal amplification. After this transient there is a slow decay
from the white spot  into the final stationary dissipative soliplasmon
state (end point of the yellow arrow). The asymptotic value for $\boldsymbol{S}(z)$
exactly coincides with the stationary spin value $\boldsymbol{S}_{a}$
calculated in the previous section. 

\begin{figure*}
\includegraphics[width=0.8\textwidth]{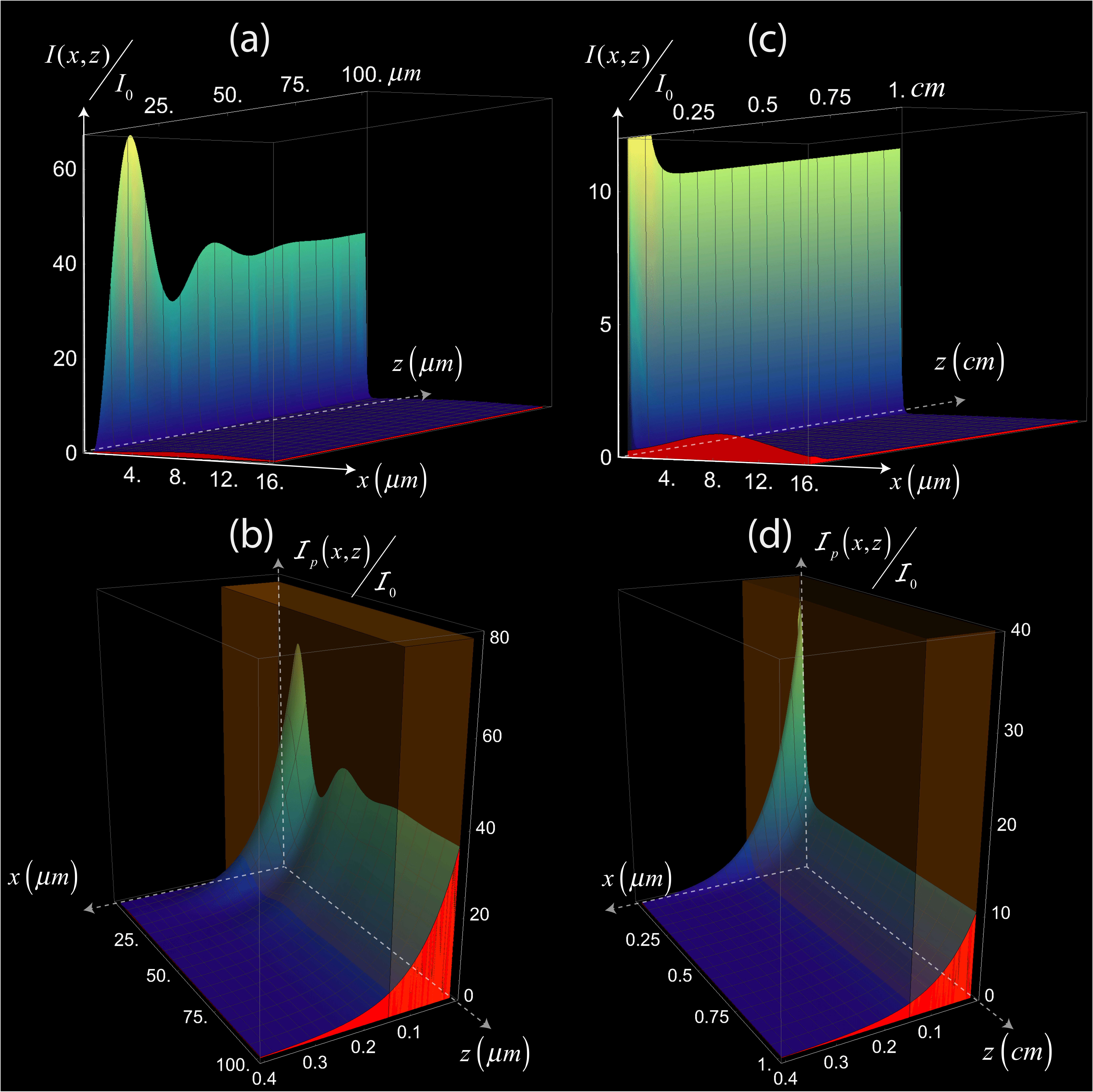}

\caption{Normalized intensity plots for the nonlinear SPP amplification process
represented in the ``Earth-like'' Poincaré sphere of Fig.~\ref{fig:Nonlinear-amplification_sphere}
for two different values of the maximum propagation length $L$. We
use normalized values given by $I/I_{0}=\left|E_{x}\right|^{2}/E_{0}^{2}$,
where $E_{0}$ is the normalization value referred to in the text.
Physical units are used for distances. Left column graphics show the
short distance behavior of the intensity for $L=20\,\mu\mathrm{m}$:
(a) intensity profile of the total soliplasmon solution; and (b) intensity
profile of the linear SPP mode excited by the soliton. Right column
shows the same evolution but for very long distances when $L=1\,cm$:
(c) full soliplasmon solution; and (d) linear SPP corresponding excited
mode.\label{fig:intensity-plots3D}}
\end{figure*}

The graphic representation in the ``Earth-like'' Poincaré sphere
is a faithful reflection of the behavior of the intensity evolution.
The intensity can be calculated from the soliplasmon \emph{ansatz}
(\ref{eq:variational_ansatz-1}) since it is given by the $z$-component
of the Poynting vector so that $I=E_{x}H_{y}^{*}\propto\left|E_{x}\right|^{2}$.
Once the spin trajectory $\boldsymbol{S}\left(z\right)$ has been
resolved, the variational parameters $C_{p}\left(z\right)$ and $C_{s}\left(z\right)$
are obtained and the soliplasmon \emph{ansatz }(\ref{eq:variational_ansatz-1})
provides an explicit solution for $E_{x}(x,z)$ and thus for the intensity
$I\left(x,z\right)$. 

Strong and fast SPP amplification in the initial propagation stage
is certainly apparent in the behavior of intensity. It occurs at typical
distances of tens of microns, as visualized in Figs.~\ref{fig:intensity-plots3D}(a)
and (b). The peak value for the intensity corresponds to a point of
maximum SPP amplification \textemdash the white spot in the sphere
in Fig.~\ref{fig:Nonlinear-amplification_sphere}, which for this
particular configuration occurs for a propagation distance of around
13 $\mu\mathrm{m}$. The nonlinear amplification of the plasmonic
mode can be more neatly seen if we use the intensity of the \emph{linear}
SPP mode associated to the soliplasmon solution at every axial position
instead of the total one. The associated SPP field is easily calculated
by projecting the soliplasmon solution (\ref{eq:variational_ansatz-1})
into a normalized SPP mode \textemdash given by the TM pair $\left(H_{py}(x),E_{px}(x)\right)$
and corresponding to the linear waveguide constituted by the MD interface\textemdash{}
at every propagation step:
\[
E_{px}^{\mathrm{proj}}(x,z)=\left[\int dxH_{py}^{*}(x)E_{x}(x,z)\right]E_{px}(x).
\]
The intensity $\mathcal{I}_{\mathrm{p}}\propto\left|E_{px}^{\mathrm{proj}}\right|^{2}$associated
to this projection can be interpreted in fact as the intensity effectively
stored into the plasmonic component of the soliplasmon mode during
propagation. So it is an optimal physical magnitude to quantify the
amplification/attenuation dynamics in the plasmonic waveguide. The
feature of maximum SPP amplification is thus apparent in Fig.~\ref{fig:intensity-plots3D}(b),
where the normalized intensity $\mathcal{I}_{\mathrm{p}}(x,z)$ is
represented up to a distance of 100 $\mu\mathrm{m}$. 

Asymptotic stabilization into a dissipative soliplasmon state is visualized
in Figs.~\ref{fig:intensity-plots3D}(c) and (d), where the propagation
of the intensity is shown for a much larger distance of $L=1\,\mathrm{cm}$.
The final profile in Fig.~\ref{fig:intensity-plots3D}(c) exactly
corresponds to the stationary soliplasmon solution obtained using
the spin model and represented by the yellow arrow in the Earth-like
Poincaré sphere (Fig.~\ref{fig:Nonlinear-amplification_sphere}).
The proximity to the north pole in the sphere indicates that the relative
weight of the plasmonic component with respect to the solitonic one
is much higher. In fact, soliton modulation is hardly visible in the
intensity profile in Fig.~\ref{fig:intensity-plots3D}(c) as soon
as the SPP component stabilizes. Besides, the whole soliplasmon solution
is asymptotically amplified. This feature is reflected in Fig.~\ref{fig:Nonlinear-amplification_sphere}
by the fact that the end of the yellow arrow lies on the atmospheric
layer. This overall amplification is smaller than for the intermediate
white point, whose norm is maximum, but it goes mostly to the SPP
component. The separate effect on the plasmonic component is clearly
seen in Fig.~\ref{fig:intensity-plots3D}(d), where we represent
the normalized projected intensity $\mathcal{I}_{\mathrm{p}}(x,z)$
up to a distance of 1 $\mathrm{cm}$. In our example, maximal amplification
of the SPP mode occurs al short distances \textemdash around 13 $\mu\mathrm{m}$\textemdash{}
but asymptotic amplification is achieved after around $200\,\mu\mathrm{m}$.
From this point on, infinitely long stationary propagation occurs
provided the system conditions are preserved.

\section{Gain-loss balance and energy flux}

\begin{figure*}
\includegraphics[width=1\textwidth]{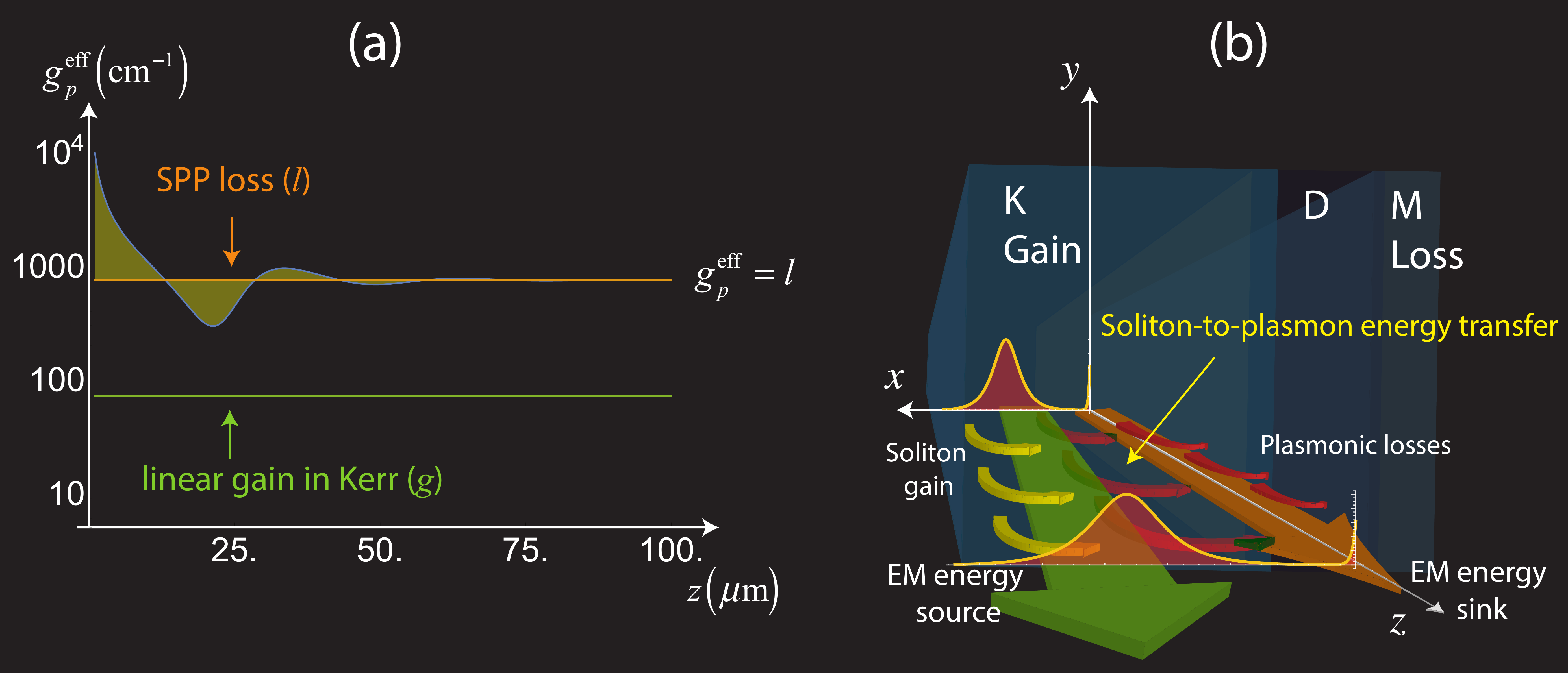}

\caption{(a) Effective gain as a function of distance showing perfect asymptotic
loss compensation \textemdash we restore physical units ($\mathrm{cm}^{-1}$)
to gain and loss coefficients and take the same configuration and
parameters as in Sections \ref{sec:Stationary-critical-gain} and
\ref{sec:Nonlinear-amplification}. (b) Graphical representation of
the typical energy flux configuration for a stationary dissipative
soliplasmon with gain in the Kerr medium compensating for SPP losses.\label{fig:gain_loss-balance+energy_flux}}

\end{figure*}

The plasmonic intensity $\mathcal{I}_{\mathrm{p}}$ introduced in
the previous section can be also used to understand the gain-loss
dynamics of the soliplasmon system. Its spatial integral $I_{p}(z)=\int dx\mathcal{I}_{p}(x,z)$
can be systematically evaluated at every axial distance once the variational
parameters $C_{p}(z)$ and $C_{s}(z)$ are determined. The complete
knowledge of the function $I_{p}(z)$ permits to introduce the concept
of \emph{effective} plasmonic gain function $g_{p}^{\mathrm{eff}}\left(z\right)$
in the following way:
\begin{equation}
\frac{dI_{p}\left(z\right)}{dz}=2\left(g_{p}^{\mathrm{eff}}\left(z\right)-l\right)I_{p}\left(z\right).\label{eq:plasmon_intensity}
\end{equation}
This form permits to understand the soliplasmon stabilization dynamics
in an analogous way as the onset of laser oscillations in an idealized
laser cavity with losses. Our equation for $I_{p}$ (\ref{eq:plasmon_intensity})
is analogous to the photon rate equation for a lossy laser cavity
in which the threshold gain \textemdash the minimum gain compensating
all cavity losses\textemdash{} is given here by the SPP loss coefficient
$l$ \cite{Milonni2010}. As in the laser cavity, also a nonlinear
dissipative system, the steady state is only achieved when gain balances
loss completely after a non trivial transient. We can see in Fig.~\ref{fig:gain_loss-balance+energy_flux}(a)
how the stabilization of the dissipative soliplasmon solution can
be neatly interpreted in terms of the balance between the effective
plasmonic gain $g_{p}^{\mathrm{eff}}$ \textemdash calculated using
Eq.(\ref{eq:plasmon_intensity})\textemdash{} and the plasmonic loss.
Indeed, the perfect gain-loss balance condition for the plasmonic
component is asymptotically fulfilled 
\begin{equation}
g_{p}^{\mathrm{eff}}=l.\label{eq:gain-loss_balance}
\end{equation}
In fact, the gain-balance condition (\ref{eq:gain-loss_balance})
defines the onset of the dissipative soliplasmon resonance. It is
remarkable that this condition can be achieved despite the nominal
linear gain in the Kerr medium $g$ is much smaller than $l$.

In order to understand the physical origin of the \emph{effective}
plasmonic distributed gain, we need to focus on the energy flux dynamics
within the soliplasmon structure. If we consider the SPP subsystem
as the one defined by the domain $\Omega_{\mathrm{MD}}$ ($x\le d$)
\textemdash containing most of the SPP amplitude, Eq.(\ref{eq:plasmon_intensity})
can be understood as the continuity equation for the energy stored
in the plasmonic subsystem. The second term in Eq.(\ref{eq:plasmon_intensity})
accounts for the rate of energy absorbed by an electromagnetic ``sink''
present in the SPP subsystem, whose origin is nothing but the losses
at the metal layer. The first term is thus naturally interpreted as
the flux of electromagnetic energy traversing the boundary of our
selected domain $\Omega_{\mathrm{MD}}$, which is given by the transverse
flux at $x=d$. Thus, this transverse flux $\Phi_{t}$ has to be proportional
to the $x-$component of the Poynting vector at this point: $\Phi_{t}(z)\propto S_{x}(d,z)$.
In the absence of gain ($g=0$) this situation is qualitatively reflected
in Fig.~\ref{fig:MDK_energy_flux}(a). Plasmonic losses act as an
electromagnetic energy sink partially compensated by the flux of energy
arising from the soliton. However, this soliton-to-plasmon energy
flux is not enough to compensate plasmonic losses so that the plasmonic
component continuously decays and eventually disappears. This qualitative
analysis is confirmed by the numerical simulations of soliplasmon
propagation using the full-vector nonlinear Maxwell's equations in
Ref.\cite{Milian2012a}. 

As we will see next, the nontrivial features of the energy flux dynamics
in our previous numerical simulation are also captured by our soliplasmon
model in a very appealing way. By construction, the energy stored
in the plasmonic subsystem can be easily evaluated using the soliplasmon
model since, as a first approximation, $I_{p}(z)=K\left|C_{p}(z)\right|^{2}$,
where $K$ is a proportionality constant By writing the variational
components in Eqs.(\ref{eq:NL_oscillator_loss_gain}) as $C_{p,s}(z)=\left|C_{p,s}(z)\right|e^{i\phi_{p,s}(z)}$
and separating real and imaginary parts, we can obtain an equation
for $d\left|C_{p}\right|/dz$. Since $dI_{p}/dz=2K\left|C_{p}\right|d\left|C_{p}\right|/dz$
we obtain immediately
\begin{equation}
\frac{dI_{p}\left(z\right)}{dz}=2\left(Kq\frac{\left|C_{p}(z)\right|}{\left|C_{s}(z)\right|}\sin\phi_{sp}(z)-l\right)I_{p}\left(z\right).\label{eq:plasmon_intensity_model}
\end{equation}
By comparing Eq.(\ref{eq:plasmon_intensity}) to Eq.(\ref{eq:plasmon_intensity_model})
and according to our previous arguments regarding the transverse energy
flux, it becomes clear that the following relations are fulfilled
\begin{equation}
\Phi_{t}(z)\propto g_{p}^{\mathrm{eff}}\left(z\right)=Kq\frac{\left|C_{p}(z)\right|}{\left|C_{s}(z)\right|}\sin\phi_{sp}(z).\label{eq:EM_flux_soliplasmon_model}
\end{equation}
This equation permits to understand the structure of the electromagnetic
energy flux in a stationary dissipative soliplasmon. For a stationary
dissipative soliplasmon the absolutes values $\left|C_{p}\right|$
as $\left|C_{s}\right|$ are $z-$independent (consequently, so is
$q$) as well as the relative phase $\phi_{sp}$. Therefore, by comparing
Eq.(\ref{eq:EM_flux_soliplasmon_model}) and the balance condition
(\ref{eq:gain-loss_balance}) we immediately recognize that, once
the modulus $\left|C_{s}\right|$ is fixed by the ``golden constraint''
(and thus also $\left|C_{p}\right|$ ), the value of the relative
phase cannot be arbitrary. In terms of the energy flux, $\phi_{sp}$
has to adjust its value in order to guarantee an energy flux that
balances plasmon losses completely. For our particular solution in
Fig.~\ref{fig:Intensity-and-phase_dissipative_soliplamons}, for
example, $\phi_{sp}\approx\pi/8$. Put it differently, in a stationary
dissipative soliplasmon the relative phase $\phi_{sp}$ determines
the exact electromagnetic energy flux flowing from the soliton to
the SPP compensating plasmon losses in an exact manner. The equilibrium
is achieved since the electromagnetic gain in the Kerr medium is perfectly
in balance with loss in the metal through the soliton-to-plasmon energy
flux, as graphically depicted in Fig.~\ref{fig:gain_loss-balance+energy_flux}(b). 

It is remarkable how the soliplasmon model explains why the existence
of a non-zero flux balance is \emph{not} achievable when there is
no gain in the Kerr medium. When $g=0$, the ``golden constraint''
as written in Eq.~(\ref{eq:golden_constraint_rewritten}) tells us
that the product $q'(C_{s})\overline{q}'(C_{s})$ has to vanish, which
implies that the only solution is the trivial one $|C_{s}|=0$. The
soliton component disappears in the soliplasmon model equations (\ref{eq:NL_oscillator_loss_gain})
leaving only a decaying plasmonic component who asymptotically vanishes:
$|C_{p}|\rightarrow0$. Thus, no nontrivial dissipative solutions
can exist.

A different situation occurs in the conservative case, when $l=g=0$.
As mentioned in Section \ref{sec:Stationary-dissipative-soliplasm},
the ``golden constraint'' disappears. In addition and according
to the balance equation (\ref{eq:gain-loss_balance}), since there
is no loss to compensate the effective plasmonic gain is zero, Thus,
the electromagnetic transverse flux $\Phi_{t}$ has to vanish, which,
according to Eq.(\ref{eq:EM_flux_soliplasmon_model}), implies that
the relative soliplasmon phase can only be either $0$ or $\pi$.
This is exactly the situation describing the conservative soliplasmons
depicted in Fig.~(\ref{fig:Conserv-soliplasmons-in-Earth-Poincare}).
Inasmuch as there is no ``golden constraint'' we have the two families
of $0-$ and $\pi-$ soliplasmon solutions parametrized by a continuous
range of values of $|C_{s}|$.

\section{Conclusions}

In this work we have established a gain/loss generalization of the
mechanism for the excitation of lossy plasmonic modes using spatial
solitons previously reported in \cite{Bliokh2009,Milian2012a}. The
key point is that this excitation mechanism occurs in the neighborhood
of a soliplasmon resonance. In this generalization we define a noteworthy
strategy to introduce gain in the system. Linear gain $g$ is introduced
in the region where the nonlinear Kerr medium is located (see Fig.
1(b).) The strategy of including gain in the Kerr medium is suggested
by the previous mechanism of nonlinear excitation of SPPs, in which
the existence of a soliplasmon resonance plays a crucial role. The
nearly resonant state formed by the SPP and the soliton \textendash the
soliplasmon resonance\textemdash{} exhibits the phenomenon of anti-crossing
\cite{Bliokh2009}, thus indicating the possibility of adiabatically
(or \emph{diabatically} if loss and gain are introduced) making transitions
between the two modes \cite{Novotny2010}. In our case, numerical
simulations of full-vector nonlinear equations have shown that the
soliton is able to excite its SPP partner in the presence of plasmonic
losses in a relatively efficient manner, increasing the SPP propagation
length \cite{Milian2012a}. For this reason, we expect that adding
a moderate gain in the Kerr medium can considerably enhance this effect.
Indeed, we have proven that by adding a small imaginary part to the
soliton propagation constant $\mu_{s}=\mu_{s}'-ik_{0}g$ it is possible
to compensate plasmonic losses \emph{completely}. With gain coefficients
$g$ significantly smaller than the plasmonic loss $l$ it is possible
to excite a soliplasmon resonance in a way that the imaginary part
of its propagation constant $\mu$ becomes exactly zero, so that the
propagation length associated to the SPP component of the coupled
system becomes infinite. 

The key point of this nonlinear mechanism of plasmonic amplification
is, however, not only the existence of a resonant interaction between
the SPP and the soliton but that this interaction is dictated by an
\emph{asymmetric }and \emph{nonlinear} evanescent coupling. It is
the fact that the two different coupling coefficients depend on the
soliton amplitude as $q,\bar{q}\propto\exp\left(-K\left|C_{s}\right|a\right)$,
which makes the existence of stationary dissipative soliplasmons \textemdash \emph{qualitatively}
and \emph{quantitatively}\textemdash{} different from what one would
obtain by assuming standard linear couplings (i.e., independent of
$\left|C_{s}\right|$.) In particular, the presence of a \emph{nonlinear}
evanescent coupling between the SPP and the soliton predicts the existence
of \emph{two} dissipative solutions \emph{always}, provided gain $g$
remains below a critical maximum value $g_{c}$. As a corollary, our
model predicts that nonlinear plasmonic amplification is critical
($g\le g_{c}$) and asymmetric ($l\gg g$). The same analysis using
a linear coupling instead leads to a completely different prediction
(see Fig.\ref{fig:golden_constraint}.) The nonlinear evanescent coupling
is a direct consequence of the soliplasmon \emph{ansatz} used to obtain
the variational equations leading to the soliplasmon model (\ref{eq:NL_oscillator_loss_gain}),
as rigorously demonstrated from nonlinear Maxwell's equations in Ref.\cite{Ferrando2013b}.
Asymmetry between the plasmon-to-soliton coupling $\bar{q}$ and the
soliton-to-plasmon coupling $q$ in realistic configurations ($\bar{q}\ll q$)
confers also peculiar properties to the soliplasmon model and, thus,
to the nonlinear mechanism of plasmonic amplification associated to
it.

The dissipative soliplasmon model presented here can be applied to
other solutions than a SPP and a spatial soliton. It can be applied
to a more general plasmonic mode of a 1D or 2D generic plasmonic waveguide
coupled to a spatially separated nonlinear dielectric mode associated
to an intricate dielectric/Kerr waveguide \cite{Ferrando2013b}. For
1D structures, for example, the plasmonic mode can be a LRSPP and
the dielectric mode can represent the nonlinear mode of a dielectric/Kerr/dielectric
waveguide. The qualitative features of the mechanism of nonlinear
plasmonic amplification mediated by dissipative soliplasmon resonances
here presented apply then to a more general set of plasmonic structures.
In this direction, the dissipative soliplasmon model (\ref{eq:NL_oscillator_loss_gain}),
or, equivalently, its spin model version (\ref{eq:spin_equations}),
describes a general behavior shared by more gain-assisted nonlinear
plasmonic guiding structures. We focussed here in the features associated
to plasmonic amplification since compensation of plasmonic losses
as been the main motivation of this work. However, the particular
nature of the soliplasmon model, with its peculiar \emph{asymmetric}
and \emph{nonlinear} evanescent coupling, makes this model certainly
much richer, as previous analysis for the particular case of symmetric
coupling show \cite{Eksioglu2012,Eksioglu2013a}. In this sense, the
use of known techniques developed to analyze dissipative solitons
in nonlinear systems \cite{Akhmediev2008a} can be applied to describe
properly the nature of all potential solutions of the soliplasmon
model beyond those find in the present work. 

Last but not the least, the soliplasmon model (\ref{eq:NL_oscillator_loss_gain})
presents interesting connections to Parity-Time (PT) symmetric optical
systems \cite{Musslimani2008,Suchkov2016a}. It can be related to
the discrete model describing the so-called nonlinear PT-invariant
dimers, represented physically by two linearly coupled active and
passive nonlinear waveguides with balanced gain and loss \cite{Suchkov2011,Zezyulin2012,Kevrekidis2013,Li2014}.
Despite the non-hermiticity of the Hamiltonian these PT-symmetric
systems admit stationary dissipative soliton solutions, in which loss
is fully balanced by gain. In our model the coupling is both asymmetric
and nonlinear and, nevertheless, stationary dissipative soliplasmons
are equally found with \emph{unbalanced} gain and loss. Interestingly,
in other structures with \emph{unbalanced} gain and loss, mathematically
represented by coupled Ginzburg-Landau equations \emph{but} with linear
and symmetric coupling, it is possible to find also stable dissipative
solitons \cite{Malomed1996,Atai1998,Nistazakis2002}. In this context,
the dissipative soliplasmon appears as a peculiar form of a dissipative
soliton, whose existence and properties rely on the asymmetric and
the nonlinear evanescent character of the plasmon-soliton coupling
near resonance. In this article we have used some of the particular
properties of the dissipative soliplasmon model aiming at optimizing
plasmonic amplification. However, many other features of the model
are still unexplored. Since nonlinear plasmonic amplification occurs
due to the mediation of the dissipative soliplasmon quasi-particle,
it is of great interest for other potential applications to deepen
in the distinguishing properties of this type of solution. 

This work was supported by the MINECO (Government of Spain) under
Grants No. TEC2010-15327, TEC2013-50416-EXP and TEC2014-53727-C2-1-R.

\appendix

\section{Variational equations\label{sec:Variational-equations}}

The variational equations describing the interaction of a linear surface
plasmon polariton (SPP) and a spatial soliton propagating in parallel
to a Metal/Dielectric/Kerr (MDK) double interface were demonstrated
in Ref.\cite{Ferrando2013b} (see Fig.\ref{fig:MDK_energy_flux}).
The linear behavior of this system is described by a permittivity
function given by:
\begin{align}
\varepsilon_{L}(x) & =\begin{cases}
\varepsilon_{p}(x) & x\le d\\
\varepsilon_{\mathrm{K}} & x>d,
\end{cases}\label{eq:linear_permittivity}
\end{align}
where $d$ is the thickness of the dielectric layer, $\varepsilon_{K}$
is the dielectric constant of the Kerr medium, and $\varepsilon_{p}(x)$
is the profile of the permittivity that defines the MD interface where
the SPP propagates, i.e., 
\begin{align}
\varepsilon_{p}(x) & \equiv\begin{cases}
\varepsilon_{m} & \mathrm{if\,\,}x\le0\\
\varepsilon_{d} & \mathrm{if\,\,}x>0,
\end{cases}\label{eq:permittivity_MD}
\end{align}
$\varepsilon_{m}$ and $\varepsilon_{d}$ being the dielectric constants
of the metal and the dielectric, respectively. In a first approach,
the metal is considered to be ideal so that $\mathrm{Im}(\varepsilon_{m})=0$. 

The resonant interaction between the spatial soliton and the SPP is
well described by a variational $TM$ solution for the electric field
$\mathbf{E}(x,z)=\left(E_{x},0,E_{z}\right)$ of the form:
\begin{equation}
E_{x}(x,z)=C_{p}(z)e_{p}(x)+C_{s}(z)\mathrm{sech}\left[\sqrt{\frac{\gamma}{2}}\left|C_{s}(z)\right|(x-a)\right],\label{eq:variational_ansatz}
\end{equation}
where 
\begin{align}
e_{p}(x) & =\varepsilon_{p}^{-1}(x)\times\begin{cases}
e^{\kappa_{m}x} & x\le0\\
e^{-\kappa_{d}x} & x>0
\end{cases}\equiv\varepsilon_{p}^{-1}(x)f_{p}(x)\label{eq:def_ep}
\end{align}
represents the linear SPP field, $k_{m}$ and $k_{d}$ being the inverse
penetration lengths in the metal and dielectric, respectively. These
quantities are explicit functions of the dielectric constants of the
metal and of the dielectric \cite{maier07,pitarke-rpp70_1a}:
\begin{align}
\kappa_{m} & =-k_{0}\varepsilon_{m}\left[-\left(\varepsilon_{m}+\varepsilon_{d}\right)\right]^{-1/2}\nonumber \\
\kappa_{d} & =k_{0}\varepsilon_{d}\left[-\left(\varepsilon_{m}+\varepsilon_{d}\right)\right]^{-1/2},\label{eq:km_kd}
\end{align}
$k_{0}$ being the vacuum wavenumber.

The amplitude of the SPP, $C_{p}(z)$, is the first complex variational
parameter. The second term in Eq.(\ref{eq:variational_ansatz}) represents
a soliton located at a distance $a$ from the the MD interface and
it has the standard form of a \emph{sech }function with an amplitude
$C_{s}(z)$, which constitutes the second variational parameter of
the model. The nonlinear coefficient $\gamma$ in the soliton functional
expression is given by $\gamma=\left(3/4\right)\varepsilon_{0}cn_{2}$,
where $n_{2}$ is the ordinary nonlinear index of the Kerr medium.
As one can see, all the non-variational parameters of the \emph{ansatz
}(\ref{eq:variational_ansatz}) are fixed. The soliton position $a$
is an input of the model as well as the rest of non-variational parameters,
which are given in terms of the physical constants defining the MDK
structure.

In this variational approach the axial component $E_{z}$ can be obtained
approximately using a transversality constraint once $E_{x}$ is determined.
The variational equations for $C_{p}$ and $C_{s}$ are obtained under
the following assumptions: \emph{(i)} propagation is quasi-stationary
(i.e., phases of variational parameters are assumed to change faster
than their modulus); \emph{(ii)} propagation is paraxial and preserves
the quasi-transverse condition ($|E_{z}|\ll|E_{x}|$); and \emph{(iii)}
the SPP-soliton coupling is small (i.e., overlapping of SPP and soliton
amplitudes in Eq.(\ref{eq:variational_ansatz}) is small.) The variational
parameters $C_{p}(z)$ and $C_{s}(z)$ define then the solution completely
through the following nonlinear model of coupled oscillators: 
\begin{eqnarray}
-i\frac{dC_{p}}{dz} & = & \mu_{p}C_{p}+q\left(C_{s}\right)C_{s}\nonumber \\
-i\frac{dC_{s}}{dz} & = & \mu_{s}\left(C_{s}\right)C_{s}+\overline{q}\left(C_{s}\right)C_{p},\label{eq:NL_oscillator_model}
\end{eqnarray}
where
\begin{align}
\mu_{p} & \equiv\frac{\left(\beta_{p}^{2}-k_{0}^{2}\varepsilon_{\mathrm{K}}\right)}{2k_{0}\varepsilon_{\mathrm{K}}^{1/2}}\nonumber \\
\mu_{s}(C_{s}) & \equiv\frac{\left(\beta_{s}^{2}-k_{0}^{2}\varepsilon_{\mathrm{K}}\right)}{2k_{0}n_{\mathrm{K}}}=\frac{k_{0}\gamma}{4\varepsilon_{\mathrm{K}}^{1/2}}|C_{s}|^{2}\label{eq:paraxial_prop_const}
\end{align}
are the paraxial propagation constants of the plasmon and soliton,
respectively. As it is well known, the plasmon non-paraxial propagation
constant $\beta_{p}$ is explicitly given in terms of the dielectric
and metal dielectric constants \cite{maier07,Pitarke2007}:
\begin{equation}
\beta_{p}=\sqrt{\frac{\varepsilon_{m}\varepsilon_{d}}{\varepsilon_{m}+\varepsilon_{d}}},\label{eq:SPP_propagation_constant}
\end{equation}
whereas the paraxial soliton propagation $\mu_{s}$ in Eqs.(\ref{eq:paraxial_prop_const}),
as expected, presents a quadratic nonlinear dependence on the soliton
variational parameter $C_{s}$ since:
\[
\beta_{s}=k_{0}\left(\varepsilon_{\mathrm{K}}+\frac{\gamma}{2}|C_{s}|^{2}\right)^{1/2}.
\]
Remarkably, this is not the only nonlinearity of this coupled oscillator
model. The variational calculation in Ref.\cite{Ferrando2013b} predicts
that the coupling coefficients in Eqs.(\ref{eq:NL_oscillator_model})
exhibit also an \emph{implicit} nonlinear dependence on the soliton
variational parameter $C_{s}$ through the overlapping integrals over
the MD and soliton domains $\Omega_{\mathrm{MD}}$ ($x\le d$) and
$\Omega_{s}$ ($x>d$): 
\begin{align}
q\left(C_{s}\right) & \equiv\frac{k_{0}}{2\varepsilon_{K}^{1/2}N_{p}}\int_{\Omega_{\mathrm{MD}}}\negthickspace\negthickspace\negthickspace\negthickspace\negthickspace\negthickspace dxf_{p}(x)\triangle\varepsilon_{s}(x)\mathrm{sech}\left[\sqrt{\frac{\gamma}{2}}\left|C_{s}\right|(x-a)\right]\nonumber \\
\bar{q}(C_{s}) & \equiv\frac{k_{0}}{2\varepsilon_{K}^{1/2}N_{s}}\int_{\Omega_{s}}\!\negthickspace\negthickspace\negthickspace dxf_{p}(x)\triangle\varepsilon_{p}(x)\mathrm{sech}\left[\sqrt{\frac{\gamma}{2}}\left|C_{s}\right|(x-a)\right]\nonumber \\
 & +\overline{q}_{V}\left(C_{s}\right)\label{eq:NL_coupling_coeff}
\end{align}
where $\triangle\varepsilon_{s}=\varepsilon_{L}-\varepsilon_{s}$
and $\triangle\varepsilon_{p}=\varepsilon_{L}-\varepsilon_{p}$ are
the local permittivity functions, which are also completely determined
by the dielectric constants of the metal, of the linear dielectric
and of the Kerr medium in the following way:
\[
\triangle\varepsilon_{p}(x)\,\,\,=\,\,\,\begin{cases}
0 & x\le d\\
\varepsilon_{\mathrm{K}}-\varepsilon_{d} & x>d
\end{cases}
\]
and
\begin{equation}
\triangle\varepsilon_{s}(x)\,\,\,=\,\,\,\begin{cases}
\varepsilon_{m}-\varepsilon_{\mathrm{K}} & x\leq0\\
\varepsilon_{d}-\varepsilon_{\mathrm{K}} & 0<x\le d\\
0 & x>d,
\end{cases}\label{eq:local_permittivity}
\end{equation}
$d$ being the width of the linear dielectric slab (see Fig.\ref{fig:MDK_energy_flux}).
The term $\overline{q}_{V}$ is the contribution to the plasmon-to-soliton
coupling of the vector term associated to the gradient of the SPP
dielectric function $\varepsilon_{p}^{-1}\nabla\varepsilon_{p}$.
There is no such a contribution for the soliton-to-plasmon coupling
$q$ because we consider the soliton to fulfill a scalar equation.
Note that, according to their definitions, the local permittivity
functions $\triangle\varepsilon_{p}$ and $\triangle\varepsilon_{s}$
are only nonzero in the soliton $\Omega_{s}$ and $\Omega_{\mathrm{MD}}$
domains, respectively. This property justifies the integration domains
in Eqs.(\ref{eq:NL_coupling_coeff}). On the other hand, the parameters
$N_{p}$ and $N_{s}$ are normalization constants, which are given
in terms of the plasmon and soliton inverse penetration lengths as:
\begin{align}
N_{p} & =\int\!\!dxf_{p}^{2}(x)\approx\frac{1}{2}\left(\kappa_{d}^{-1}+\kappa_{m}^{-1}\right)\nonumber \\
N_{s} & =\int\!\!dx\,\mathrm{sech^{2}}\left[\sqrt{\frac{\gamma}{2}}\left|C_{s}\right|(x-a)\right]\approx2\kappa_{s}^{-1}.\label{eq:norms}
\end{align}
Note that since the soliton penetration length also depends on the
shape of the soliton profile, there is an extra dependence on the
soliton variational parameter given by $\kappa_{s}=\left(\gamma/2\right)^{1/2}k_{0}|C_{s}|$.
This fact along with the expressions for the coupling coefficients
(\ref{eq:NL_coupling_coeff}) clearly point out both the asymmetry
and the nontrivial nonlinear nature of the resonant soliplasmon coupling. 

The variational model given by the nonlinear coupled oscillator system
(\ref{eq:NL_oscillator_model}) is a model with no free parameters.
All coefficients entering these equations are given in last instance
in terms of the optical coefficients that characterize the MDK structure
\textemdash i.e., $\varepsilon_{m}$, $\varepsilon_{d}$, $\varepsilon_{K}$,
the width $d$, and the nonlinear index $n_{2}$\textemdash{} along
with the soliton position $a$. In this sense, within the regime of
validity of the approximations used to obtain it, the variational
model possesses the same predictive power as the original Maxwell's
equations from which it was derived.

The inclusion of metal losses ($\mathrm{Im}(\varepsilon_{m})\ne0$)
does not change the equations for the SPP. They remain valid assuming
now that $\varepsilon_{m}$ is complex \cite{maier07}. For this reason
the structure of all the previous equations does not change. However,
now the plasmon function $f_{p}$ is complex since both $\kappa_{m}$
and $\kappa_{d}$ are, according to Eq.(\ref{eq:km_kd}). The SPP
propagation constant (\ref{eq:SPP_propagation_constant}) becomes
complex as well:

\[
\beta_{p}^{2}=\beta'{}_{p}^{2}+i\beta''{}_{p}^{2}.
\]
Consequently, and according to (\ref{eq:paraxial_prop_const}), the
paraxial propagation constant becomes also complex: $\mu_{p}=\mu'{}_{p}+i\mu''{}_{p}$.
The same occurs to the coupling coefficients in (\ref{eq:NL_coupling_coeff}),
which now can be written explicitly separating their real and imaginary
parts as:
\begin{align*}
q & =q_{m}'+i\delta q''_{m}\\
\overline{q} & =\overline{q}_{m}'+i\delta\overline{q}''_{m}.
\end{align*}
We use the subindex $m$ to indicate that the origin of these terms
is the existence of metal losses, so that $\delta q''_{m},\delta\overline{q}''_{m}\rightarrow0$
when $\mathrm{Im}\left(\varepsilon_{m}\right)\rightarrow0$. 

\section{The Earth-like Poincaré sphere}

The Poincaré sphere (or Bloch sphere) is a standard representation
of a density matrix, which is by construction Hermitian and positive
definite ($\det\rho\ge0$) \cite{Mandel1995a}. The general construction
followed in Section~ \ref{sec:Spin-model} to write arbitrary Hermitian
matrices in the $\left\{ \tau_{0},\boldsymbol{\tau}\right\} $ basis
determines that a generic density matrix $\rho$ is represented by
a 4 dimensional vector $\left(S_{0},\boldsymbol{S}\right)$ verifying
the inequality $|\boldsymbol{S}|\le S_{0}$. This inequality defines
a sphere of radius $S_{0}$ (the Poincaré or Bloch sphere) in the
3D space of spin components. A pure state represented by a density
matrix $\rho=\left|C\right\rangle \left\langle C\right|$ is the only
type of state that fulfills the equality $|\boldsymbol{S}|=S_{0}$
having the explicit representation in terms of the components of the
complex vector $C$ given in Eq.(\ref{eq:spin_vector}). Pure states
correspond then to vectors lying on the surface of the Poincaré sphere.
Non-pure states are represented by spin vectors inside the Poincaré
sphere.

When we are considering the conservative case, in which we set $\sigma=0$
in the spin model evolution equation (\ref{eq:spin_equations}), $|\boldsymbol{S}|$
is conserved and, therefore, all the allowed trajectories of the spin
$\boldsymbol{S}(z)$ occurs on the surface of the Poincaré sphere.
In the linear case, the spin vector $\boldsymbol{S}(z)$ experiments
a precession around the constant effective magnetic field $\boldsymbol{\Omega}$.
In the nonlinear case, since $\boldsymbol{\Omega}$ depends on $\boldsymbol{S}$,
the magnetic field can also evolve in $z$. However, no matter the
evolution is linear or nonlinear, the conservative character of the
dynamics preserves the norm of the spin and, therefore, the condition
that $\boldsymbol{S}$ has to move on the surface of the Poincaré
sphere.

When the spin system is non-conservative, we are in the general case
described by the spin evolution equations (\ref{eq:spin_equations})
with $\sigma\neq0$. Now, the first equation for $dS_{0}/dz$ clearly
indicates that the evolution of the spin vector do not preserve its
norm. As a consequence, evolution can take the spin vector out of
the surface of the Poincaré sphere. In the general case it is possible
that $dS_{0}/dz$ is negative or positive so we can find situations
in which $|\boldsymbol{S}|<S_{0}$ or $|\boldsymbol{S}|>S_{0}$. Consequently,
in the non-conservative case we expect that the trajectories of the
spin vectors move all over the space, inside and outside the Poincaré
sphere. 

However, the expression that we found for the zero component of a
stationary dissipative soliplasmon (\ref{eq:S0}) tells us that we
can find solutions with an infinitely large value of $S_{0}$ (when
$n_{3}\rightarrow1$). The existence domain of $S_{0}$ is then the
whole positive real axis $[0,\infty[$. Since the visualization of
spin vectors with large norm can be difficult using the standard Poincaré
sphere, we introduce a modified version of this representation. Basically,
we introduce a new function that maps the infinite $S_{0}$ domain
into the finite domain $[0,\varphi]$, where $\varphi$ is the \emph{golden
ratio}. The choice of $\varphi$ is certainly arbitrary. It is chosen
because it provides a good proportion between the spheres radii for
an equilibrated visualization of the inner and outer sphere:
\begin{align}
\overline{S}_{0}(z) & =\frac{2\varphi}{\pi}\arctan\left[\frac{S_{0}(z)}{\widetilde{S}}\tan\left(\frac{\pi}{2\varphi}\right)\right],\label{eq:S0_finite_map}\\
\bar{\boldsymbol{S}}(z) & =\overline{S}_{0}(z)\frac{\boldsymbol{S}(z)}{|\boldsymbol{S}(z)|},
\end{align}
where $\widetilde{S}$ is some reference value. The previous mapping
has the following properties:
\begin{eqnarray*}
\overline{S}_{0} & \rightarrow & 0\,\,\,\,\,\mathrm{when}\,\,\,\,\,S_{0}\rightarrow0\\
\overline{S}_{0} & \rightarrow & 1\,\,\,\,\,\mathrm{when}\,\,\,\,\,S_{0}\rightarrow\widetilde{S}\\
\overline{S}_{0} & \rightarrow & \varphi\,\,\,\,\,\mathrm{when}\,\,\,\,\,S_{0}\rightarrow\infty.
\end{eqnarray*}
In fact, the particular form of the mapping (\ref{eq:S0_finite_map})
is to some extent arbitrary since other mapping function could be
used provided it satisfied the previous requirements together with
the properties of analyticity and monotonicity in the $[0,\infty[$
domain. The advantage of such a mapping is that now all spin vectors
generated by the evolution described by the non-conservative spin
equations (\ref{eq:spin_equations}) fit always in an extended Poincaré
sphere of radius $R=\varphi$. Since $\overline{S}_{0}$ is a monotonically
increasing function it is also true that 
\begin{align*}
|\boldsymbol{S}|<\widetilde{S} & \Rightarrow0\le\overline{S}_{0}<1\\
|\boldsymbol{S}|=\widetilde{S} & \Rightarrow\overline{S}_{0}=1\\
|\boldsymbol{S}|>\widetilde{S} & \Rightarrow1<\overline{S}_{0}\le\varphi.
\end{align*}
According to these properties, the function (\ref{eq:S0_finite_map})
maps the inner part and the surface of a Poincaré sphere (of radius
$R=\widetilde{S}$) into a normalized to unit Poincaré sphere. However,
the infinite outer part is mapped into a finite shell comprised between
the unit sphere and an outer sphere of radius equals to the \emph{golden
ratio} $\varphi$. 

For example, if we chose to analyze the spin motion in a conservative
regime in which $\sigma=0$ and thus $|\boldsymbol{S}|=S_{0}^{c}$
is conserved, we would select $\widetilde{S}=S_{0}^{c}$ and therefore
all the motion would happen on the surface of the unit sphere. If
we now switched on the non-conservative terms in Eq.(\ref{eq:spin_equations})
by letting $\sigma\neq0$, the motion would cease to be on the surface
and either it would ``take off'' into the outer shell (if $dS_{0}/dz>0$)
or it would ``dig into'' the inner one ($dS_{0}/dz<0$). In this
framework, the outer shell represents the set of spin states with
overall gain (larger norm) with respect their counterparts with $\sigma=0$.
Equivalently, the inner shell constitute the set of states with overall
loss (smaller norm) with respect the same spin states with $\sigma=0$.
The physical picture of this extended version of the standard Poincaré
sphere is similar to an Earth-like system, in which conservative motion
occurs on the Earth surface, lossy evolution takes place under the
Earth surface and evolution with gain materializes into the Earth
atmosphere. For this reason, we refer to this form of representing
the solutions of the non-conservative spin model (\ref{eq:spin_equations})
as the Earth-like Poincaré sphere.

\section{Propagation constant of a stationary dissipative soliplasmon}

In the framework of our spin model, a fully determined solution of
a stationary dissipative soliplasmon is given by the spin equations
(\ref{eq:explicit_sol_n}) and (\ref{eq:S0}). Apparently, however,
the information about the propagation constant of the original stationary
state $C=\left(C_{p},C_{s}\right)^{\top}$ seems to be lost in the
process of defining the equivalent spin model. Indeed, this essential
parameter does not appear explicitly neither in the dynamical spin
equations (\ref{eq:spin_equations}) nor in the solutions for stationary
spin states (\ref{eq:stationary_sol_n}). On the other hand, the general
form of the spin vector $\boldsymbol{S}$ in terms of the components
of the $C$ vector (\ref{eq:spin_vector}) only determines the modulus
$|C_{p}|$, $|C_{s}|$ of the plasmon and soliton components as well
as their relative phase $\phi_{sp}=\phi_{s}-\phi_{p}$ but not the
individual phase of each component. In this way, given the spin vector
of a stationary solution, we can know the stationary state $C$ up
to a global phase, which in principle could depend on $z$ in an arbitrary
form:
\begin{align}
C(z) & =e^{i\phi_{p}(z)}\left[\begin{array}{c}
|C_{p}(0)|\\
|C_{s}(0)|e^{i\phi_{sp}(0)}
\end{array}\right]=e^{i\phi_{p}(z)}\widetilde{C},\label{eq:stationary_solution_C}
\end{align}
where $\widetilde{C}$ is the 2D complex vector that can be univocally
constructed from the values of the stationary spin solution.

Despite this fact, the propagation constant of the stationary dissipative
soliplasmon is also univocally defined. According to our process of
constructing our dissipative nonlinear solution, once we determine
\textendash if extant\textemdash{} the specific value of $|C_{s}|$
that satisfies the ``golden constraint'', all components of the
original nonlinear Hamiltonian $H$ (\ref{eq:hamiltonian}) are fixed.
This is so because $q$, $\overline{q}$ and $\mu_{s}$ are functions
of $|C_{s}|$ whereas $\mu_{p}$ is a constant. According to the original
Hamiltonian equation for the vector $C$ (\ref{eq:hamiltonian}),
the stationary solution (\ref{eq:stationary_solution_C}) would fulfill
this equation in the following way,
\begin{align*}
\frac{d\phi_{p}}{dz}\widetilde{C} & =H\widetilde{C},
\end{align*}
where the matrix $H$ and the vector $\widetilde{C}$ are both independent
of $z$ and fully determined by the stationary spin solution. The
solution of the equation above is thus 
\begin{align*}
\phi_{p}(z) & =\mu z,
\end{align*}
where
\[
\mu=\frac{\widetilde{C}^{\dagger}H\widetilde{C}}{\widetilde{C}^{\dagger}\widetilde{C}}=\frac{\mathrm{Tr}\left(\widetilde{\rho}H\right)}{\mathrm{Tr}\left(\widetilde{\rho}\right)},
\]
in which $\widetilde{\rho}=\widetilde{C}\widetilde{C}^{\dagger}$
is the density matrix associated to the stationary spin solution.
We can rewrite the previous equation in terms of the real four dimensional
vectors $\Omega$ , $\sigma$ and $S$ following the procedure in
Section~\ref{sec:Spin-model}.
\begin{align*}
\mu & =\left[\mathrm{Tr}\left(\Pi\widetilde{\rho}\right)+i\mathrm{Tr}\left(\Sigma\widetilde{\rho}\right)\right]/\mathrm{Tr}\left(\widetilde{\rho}\right)\\
 & =\frac{1}{2}\left(\Omega_{0}+\boldsymbol{\Omega}\cdot\boldsymbol{n}\right)+i\frac{1}{2}\left(\sigma_{0}+\boldsymbol{\sigma}\cdot\boldsymbol{n}\right).
\end{align*}
However, for stationary spin solutions the imaginary part vanishes.
The reason is that the imaginary part of $\mu$ is proportional to
$dS_{0}/dz$ since, according to the first of the spin equations (\ref{eq:spin_equations}),
we have
\[
\mathrm{Im}\left(\mu\right)=\frac{1}{2}\left(\sigma_{0}+\boldsymbol{\sigma}\cdot\boldsymbol{n}\right)=-\frac{1}{S_{0}}\frac{dS_{0}}{dz}.
\]
Inasmuch stationary solutions have invariant norm they fulfill the
condition $dS_{0}/dz=0$ and, therefore, $\mathrm{Im}\left(\mu\right)\sim dS_{0}/dz=0$. 

In summary, we have seen that the propagation constant of a stationary
dissipative soliplasmon is real and it is univocally defined by the
components of the stationary spin solution associated to it as:
\begin{align*}
\mu & =\frac{1}{2}\left[\Omega_{0}+\boldsymbol{\Omega}\cdot\boldsymbol{n}\right].
\end{align*}

\bibliographystyle{/Users/aferrand/Documents/Carpeta_Magica/latex_styles/revtex4-1/bibtex/bst/revtex/apsrev4-1}
\bibliography{/Users/aferrand/Documents/Carpeta_Magica/Investigacio/linies_de_treball/plasmonica/contribucions_nostres/articles/esborranys/dissipative_soliplasmons/bibtex/dissipative_soliplasmons}

\end{document}